\documentclass[9pt,twocolumn,twoside
]{arxiv-preprint}
\templatetype{arxiv_main_SI}
\usepackage{amsmath}
\usepackage{amssymb}
\usepackage{amsbsy} 
\usepackage{float,placeins}
\usepackage{xcolor, soul}
\usepackage{graphicx}
\usepackage{bm}

\DeclareCaptionFormat{SIcaption}{\normalfont\sffamily\fontsize{9}{9}\selectfont#1#2#3}
\newcommand{\rp}{\mathbf{r}}
\newcommand{\bp}{\mathbf{b}}

\title{Transcription-induced active forces suppress chromatin motion}

\author[a]{Sucheol Shin}
\author[a,b]{Guang Shi}
\author[c]{Hyun Woo Cho}
\author[a,d]{D. Thirumalai}

\affil[a]{Department of Chemistry, The University of Texas at Austin, Texas 78712, USA}
\affil[b]{Department of Materials Science, University of Illinois, Urbana, Illinois 61801, USA}
\affil[c]{Department of Fine Chemistry and Center for Functional Biomaterials, Seoul National University of Science and Technology, Seoul 01811, Republic of Korea}
\affil[d]{Department of physics, The University of Texas at Austin, Texas 78712, USA}

\leadauthor{Shin} 

\significancestatement{
In order to explain a physically counter-intuitive experimental finding that chromatin mobility is reduced during transcription, we introduced a polymer  model for interphase chromosome  that accounts for RNA polymerase (RNAP) induced active force. The simulations show that, over a range of active forces, the mobility of the gene-rich loci is suppressed. Outside this range, chromosomes are compact and exhibit glass-like dynamics. Our study, which accounts for the experimental observations, leads to  a novel and testable mechanism of how transcription could  shape the coexistence of fluid- and solid-like properties within chromosomes. 
}

\authorcontributions{S.S. and D.T. designed research; S.S., G.S, and D.T. performed research; S.S., G.S, H.W.C, and D.T. analyzed data, and S.S., G.S, and D.T. wrote the paper}
\authordeclaration{The authors declare no conflict of interest.}

\begin{abstract}

The organization of interphase chromosomes in a number of species is starting to emerge thanks to advances in a variety of experimental techniques. However, much less is known about the dynamics, especially in the functional states of chromatin. Some experiments have shown that the mobility of individual loci in human interphase chromosome decreases during transcription, and increases upon inhibiting transcription. This is a counter-intuitive  finding because it is thought that the active mechanical force ($F$) on the order of ten pico-newtons, generated by RNA polymerase II (RNAPII) that is presumably transmitted to the gene-rich region of the chromatin, would render it more open, thus enhancing the mobility. Inspired by these observations, we developed a minimal active copolymer model for interphase chromosomes to investigate how $F$ affects the dynamical properties of chromatin. The movements of the loci in the gene-rich region are suppressed in an intermediate range of $F$, and are enhanced at small $F$ values, which has also been observed in experiments. 
In the intermediate $F$, the bond length between consecutive loci increases, becoming commensurate with the distance at the minimum of the attractive interaction between non-bonded loci. This results in a transient disorder-to-order transition, leading to the decreased mobility during transcription.  
Strikingly, the $F$-dependent change in the locus dynamics preserves the organization of the chromosome at $F=0$. 
Transient ordering of the loci, which is not found in the polymers with random epigenetic profiles, in the gene-rich region might be a plausible mechanism for nucleating a dynamic network involving transcription factors, RNAPII, and chromatin.  
 
\end{abstract}

\dates{This manuscript was compiled on \today}
\doi{DOI to be provided by publisher}

\begin{document}

\maketitle
\thispagestyle{firststyle}
\ifthenelse{\boolean{shortarticle}}{\ifthenelse{\boolean{singlecolumn}}{\abscontentformatted}{\abscontent}}{}

Advances in experimental techniques \cite{Lieberman-aiden2009,Bintu2018a} have elucidated the organizational details of chromosomes, thus deepening our understandings of how gene regulation is connected to chromatin structure \cite{Finn2019b}.
Relatively less is known about the dynamics of the densely packed interphase chromosomes in the cell nucleus. 
Experimental and theoretical studies have shown that the locus dynamics is heterogeneous, exhibiting sub-diffusive behavior \cite{Lucas2014,Bronshtein2015,Liu2018,Ashwin2019,Shaban2020}, which is consistent with models that predict glass-like dynamics \cite{Kang2015b,Shi2018a,DiPierro2018a}.  However, it is challenging to understand the dynamic nature of chromosomes that governs the complex subnuclear processes, such as gene transcription. 

\begin{figure*}[h!]
\centering
\includegraphics[width = 6.8 in]{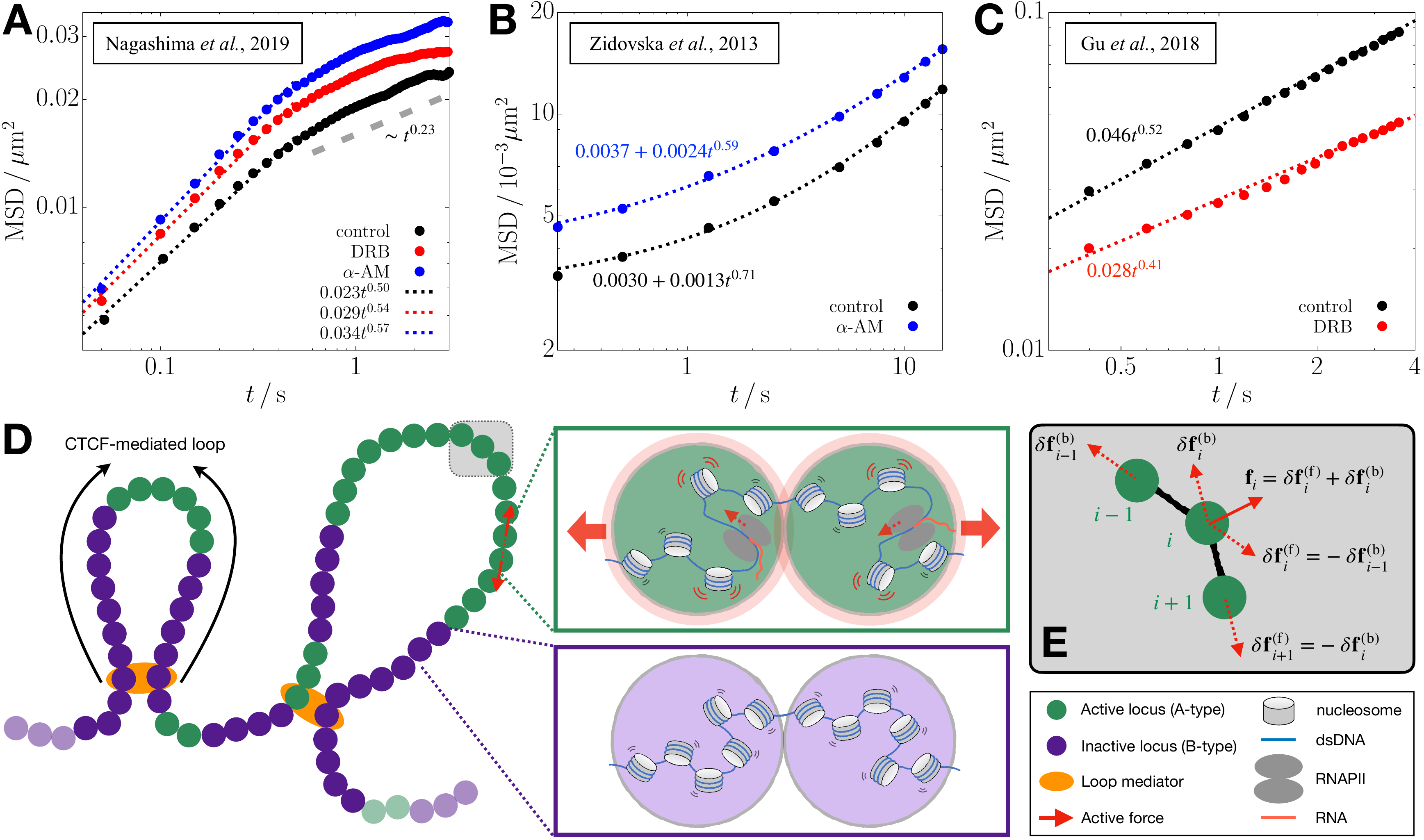}
\caption{
Comparison of chromatin mobility changes in different experiments of transcription inhibition (top) and the schematic depiction of the polymer model used in the current study (bottom). 
(A--C) Time-dependent MSD data from the experiments of Refs.~\citenum{Nagashima2019} (A), \citenum{Zidovska2013} (B), and \citenum{Gu2018} (C) are plotted on a log scale. The data for the nuclei treated with DRB and $\alpha$-amanitin ($\alpha$-AM) are shown in red and blue circles, respectively, whereas those for the control (untreated nuclei) are in black circles. The dotted lines are the best fits to the data in the given time range ($t < 0.5\,\mathrm{s}$ for panel A). The gray dashed line in panel A shows the scaling of MSD for $t > 0.5\,\mathrm{s}$ suggestive of the constrained motion of nucleosomes.
(D) Chromosome is modeled as a copolymer chain consisting of active/inactive (green/purple-colored) loci. Specific locus pairs are connected by loop mediators (orange color). We apply active forces to all the gene-rich A-type loci in the form of an extensile force dipole (red arrows) along each bond vector between the A-type loci. 
The green and purple boxes illustrate the microscopic origin of the dipolar extensile active forces. Compared to the inactive B-type loci (bottom), the A-type loci include the nucleosomes in a more dynamic state along with the actively transcribing RNAPII, such that their effective excluded volumes increase modestly, as represented by the light-red shade. The enhanced volume exclusion gives rise to  repulsion between the consecutive A-type loci and the increase of the bond distance.
(E) Illustration for how active forces are imposed on the $i^\text{th}$ locus and its bonded loci, where the solid arrow is the net force from the force dipoles (dashed arrows). 
}
\label{fig:exp_model}
\end{figure*}

The link between transcriptional activity and changes in chromosome dynamics is important in understanding the dynamics of chromosomes in distinct cell types and states \cite{Nozaki2017, Nagashima2019}. 
It is reasonable to expect that transcription of an active gene-rich region could make it more expanded and dynamic \cite{Gu2018,Tortora2020}. 
However, active RNA polymerase (RNAP) II suppressed the movement of individual loci in chromatin \cite{Zidovska2013,Nagashima2019}. 
Let us first summarize the key experimental results \cite{Nagashima2019,Zidovska2013}, which inspired our study: (1) By imaging the motion of individual nucleosomes (monomers in the chromosomes) in live human cells, it was shown that the mean square displacements (MSDs) of the nucleosomes during active transcription are constrained (see Fig.~\ref{fig:exp_model}A). 
(2) When the cells were treated with $\alpha$-amanitin ($\alpha$-AM) or 5,6-Dichloro-1-$\beta$-D-ribofuranosylbenzimidazole (DRB), transcription inhibitors which selectively block the translocation of RNAPII  \cite{Brueckner2008, Bensaude2011}, the mobilities of the nucleosomes were enhanced (Figs.~\ref{fig:exp_model}A--\ref{fig:exp_model}B). 
This finding is counter-intuitive because the translocation generates mechanical forces \cite{Yin1995,Wang1998} that should render the chromatin more dynamic.
(3) The enhanced motion was restricted only to gene-rich loci (euchromatin) that are predominantly localized in the cell interior \cite{Xing1995,Croft1999} whereas the dynamics of gene-poor heterochromatin, found in the periphery \cite{Xing1995,Croft1999,Reddy2008}, is unaffected by transcription. 
Based on these observations, it was hypothesized that RNAPs and other protein complexes, facilitating transcription, transiently stabilize chromatin by forming dynamic clusters \cite{Hnisz2017,Cho2018,Chong2018,Sabari2018,Hsieh2020}, whose structural characteristics are unknown.
This hypothesis has been challenged because the inhibition mainly leads to stalling of RNAPs bound to chromatin \cite{Brueckner2008,Bensaude2011}.
Moreover, transcriptional inhibition does not significantly alter the higher-order structures of chromosomes \cite{Hsieh2020,Jiang2020}. 
These observations raise the question: Is there a physical explanation for the increased  chromatin dynamics upon  inhibition of transcription and a decrease during transcription? We provide a plausible answer to this question by using extensive simulations of  a simple minimal active copolymer model, which captures the organization of chromosomes in the absence of active forces. 

In contrast to the experimental findings discussed above, it has been shown that under certain conditions \cite{Gu2018} the mobility of gene loci increases in the presence of transcription. In other words, the mobility decreases when transcription is inhibited (see Fig.~\ref{fig:exp_model}C). 
 This intuitive result may be ratinoalized by assuming that transcription exerts active forces although differences between the results in Figs.~\ref{fig:exp_model}C and \ref{fig:exp_model}A are hard to explain.

Using the experimental results as a backdrop, we theorize that RNAPII exerts active force \cite{Yin1995,Wang1998} in a vectorial manner on the active loci.
We then examine the effects of active force on the organization and dynamics of chromosome using the Chromosome Copolymer Model (CCM) \cite{Shi2018a}.
In the absence of active forces, the CCM faithfully captures the experimental results, showing microphase separation between euchromatin (A-type loci) and heterochromatin (B-type loci) on large length scale and formation of substructures on a smaller length scale in interphase chromosomes.
The CCM leads to a condensed polymer, exhibiting glass-like dynamics in the absence of activity, in agreement with experiments \cite{Lucas2014,Bronshtein2015,Shi2018a,Strickfaden2020}.
Brownian dynamics simulations of a single chromosome in the presence of active forces, produces the following key results: (i) In accord with experiments \cite{Nagashima2019}, 
the dynamics of the active loci, measured using the MSD, is suppressed upon application of the active force. 
Despite the use of minimal model, our simulations qualitatively capture the increase in the MSD for the transcriptionally inactive case relative to the active case.
(ii) The decrease in the mobility of the A loci occurs only over a finite range of activity level. 
Surprisingly, in this range the segregated A loci undergo a transient disorder-to-order transition whereas the B loci remain fluid-like.   (iii) Chromosome in which the epigenetic (A/B) profile is random cannot capture the experimental observations, implying that the sequence inherent in the chromosomes plays a vital role.

\section*{Model}

\subsection*{Chromosome Copolymer Model} 
We model an interphase chromosome as a flexible self-avoiding copolymer (Fig.~\ref{fig:exp_model}D), whose potential energy function is given in Methods.
Non-adjacent pairs of loci are subject to favorable interactions, modeled by the Lennard-Jones (LJ) potential (Eq.~\ref{eq:LJ}), depending on the locus type.
The relative strengths of interactions between the loci is constrained by the Flory-Huggins theory \cite{Huggins1941,Flory1941}, which ensures microphase separation between the A and B loci---an important organization principle of interphase chromosomes. 
Certain locus pairs, separated by varying genomic distances, are linked to each other, representing the chromatin loops mediated by CCCTC-binding factors (CTCFs) \cite{Rao2014}.
In this study, we simulate a 4.8-Mb segment of human chromosome 5 (Chr5: 145.87--150.67 Mb) using $N=\text{4,000}$ loci, that is, 1.2 kb $\sim$ 6 nucleosomes per locus. 
The A- to B-type ratio is $N_\text{A}/N_\text{B} =982/3018\approx1/3$. 
See Methods for the details about assignment of the locus type and the loop anchors in the CCM polymer.

In our model, the loop topology does not change with time, unlike polymer models, which examined the consequences of dynamic extrusion of loops on interphase and mitotic structures \cite{Fudenberg2016,Nuebler2018,Dey2023}.
We simulate a given chromosome region up to $\sim$10 seconds (see Methods for details) that is much shorter than the lifetime of the CTCF loops (15--30 mins) \cite{Gabriele2022,Mach2022}. Furthermore, the loop extrusion rate is $\sim$0.5--2 kb/s \cite{Kim2019,Golfier2020}, which also would not significantly alter the topology of loops on the simulation time scales.  Therefore, the assumption that loops are static is reasonable.
We discuss, in a later section, the potential effects of the removal of loops along with the implications (see Discussion)

\subsection*{Active forces in transcriptional activity} 
Previous theoretical studies \cite{Ghosh2014,Eisenstecken2016,Sakaue2017,Bianco2018,Put2019,Foglino2019,Smrek2020,Locatelli2021,Ghosh2022} have considered different forms of active forces on homopolymers and melts of polymer rings.  
Potential connection to chromatin has been proposed using polymer models in the presence of active forces \cite{Ganai2014,Liu2018,Saintillan2018,Liu2021,Mahajan2022a,Jiang2022,Goychuk2023}. However,  these have not  accounted for the counter-intuitive finding from the experiment of transcription inhibition \cite{Nagashima2019}. 
Here we applied active forces on the CCM polymer chain to mimic the force generated by the actively transcribing RNAPII during transcription elongation \cite{Yin1995,Wang1998}. 

In our model, active forces act along each bond vector between the A-type loci in an extensile manner, ensuring momentum conservation (see Fig.~\ref{fig:exp_model}D).
The force, associated with the translocation of actively transcribing RNAPII \cite{Yin1995,Wang1998}, should result in the increased volume exclusion between the bonded A loci, as shown in the green box in Fig.~\ref{fig:exp_model}D.
We reason that the dipolar extensile active force, with local momentum conservation, models the effective repulsion between the active loci in the coarse-grained level (see SI Appendix, Sec.~1 for more details about the biological relevance of the force).
Along the A-A bond vector, $\mathbf{b}_i = \mathbf{r}_{i+1}-\mathbf{r}_{i}$ ($\mathbf{r}_{i}$ is the position of the $i^\text{th}$ locus), force, $\delta \mathbf{f}_{i+1}^\mathrm{(f)} = f_0\mathbf{\hat{b}}_i$, acts on the $(i+1)^\text{th}$ locus in the forward direction ($f_0$ is the force magnitude and $\mathbf{\hat{b}}_i = \mathbf{b}_i/|\mathbf{b}_i|$), and $\delta \mathbf{f}_{i}^\mathrm{(b)}=- \delta \mathbf{f}_{i+1}^\mathrm{(f)}$ is exerted on the $i^\text{th}$ locus in the opposite direction (Fig.~\ref{fig:exp_model}E).
If the $(i-1)^\text{th}$ locus is also A-type, another pair of active force is exerted along the bond vector, $\mathbf{b}_{i-1} = \mathbf{r}_{i}-\mathbf{r}_{i-1}$, and thus the net active force on the $i^\text{th}$ locus is given by $\mathbf{f}_i = \delta \mathbf{f}_i^\mathrm{(f)}+\delta \mathbf{f}_i^\mathrm{(b)} = f_0 (\mathbf{\hat{b}}_{i-1} -\mathbf{\hat{b}}_i)$ (solid arrow in Fig.~\ref{fig:exp_model}E).
We simulate the CCM chain dynamics by numerically solving the overdamped Langevin equation, given in Eq.~\ref{eq:bd}, which includes the active forces (see Methods for simulation details).
We use the dimensionless parameter, $F \equiv f_0 \sigma/k_B T$, as a measure of the force magnitude, where  $\sigma$ is the diameter of a single locus, $k_B$ is the Boltzmann constant, and $T$ is the temperature. 

In the simulations, we apply the active forces of given magnitude $F$ to all the A-type loci at each time step. 
Our choice of the locus resolution (1.2 kb) is sufficiently small to distinguish between the gene and the intergenic loci which are assigned A- and B-types, respectively (note that the average human gene size is $\sim$27 kb \cite{Venter2001}). 
Here we assume that all the genes in the simulated chromosome region are co-activated, and experience the same magnitude of forces during our simulation. 
In a later section, we discuss the potential effects of changing the gene activations and the corresponding force exertion (see ``Connection to other experiments'' in Discussion).

On each gene body, there are multiple active RNAPII complexes during a transcriptional burst, the time in which transcripts are generated \cite{Hager2009,Rodriguez2020}.  
For a human gene, the number of RNAPII or transcribed RNAs is observed to be up to 5 during the burst interval of 10--20 minutes, which seems to be conserved across other genes \cite{Wan2021}. 
Hence, the average linear density of active RNAPII in each gene during the burst can be estimated as $\sim0.01$--0.05 RNAPII per kb. 
On the other hand, RNAPII clusters observed in live-cell imaging experiments \cite{Cisse2013,Cho2018} have the volume density of $\sim 10^4$ RNAPII per $\mu\text{m}^3$, which corresponds to $\sim$ 1 RNAPII per kb based upon genome density in the nucleus. 
Although the clusters are closely related to transcriptional activity \cite{Cho2018,Du2024}, it is unclear what fraction of RNAPII in the clusters is activated and engaged in the transcription elongation of genes. 
Moreover, it is unknown what kind of forces are generated on mammalian chromatin during transcription, which is crowded with RNAPII and other protein factors. 
Here we explore the dynamics of the model chromatin subject to varying active forces. 

In the following, we first present the simulation results where the active forces were applied on all the A-type loci. 
In the Discussion the section, we compare the main results with those from the simulations with lower force density (see ``Effect of active force density'').

\section*{Results}

\subsection*{MSD comparison}
\begin{figure}[h!]
\centering
\includegraphics[width = 3.4 in]{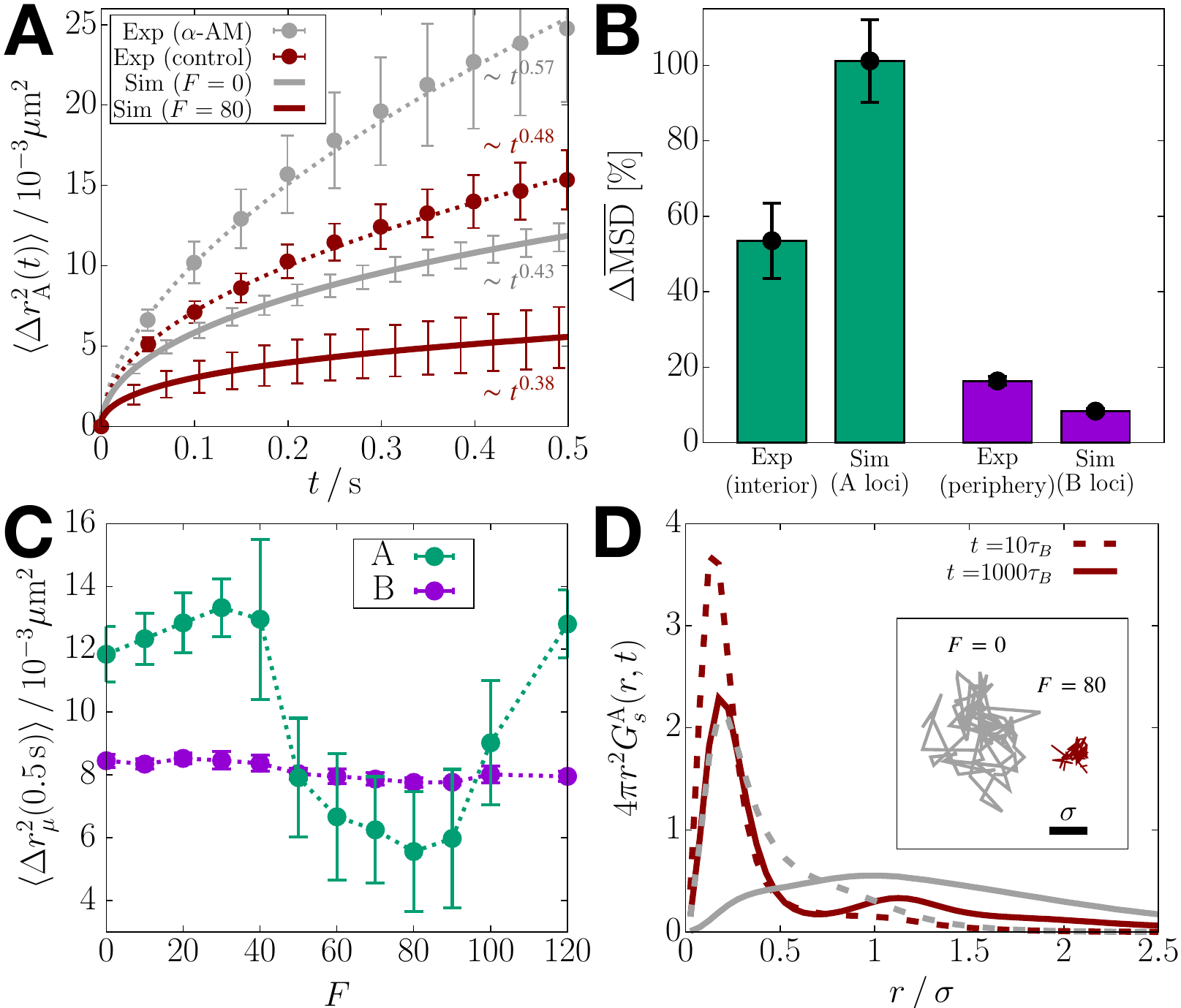}
\caption{
Transcription-induced active forces reduce the mobilities of euchromatin loci. 
(A) $\langle{\Delta r^2_\text{A}}(t)\rangle$ from simulations with $F=0$ and $F=80$ (solid lines), compared with the euchromatin MSD from the experiment that inhibits transcription using $\alpha$-AM \cite{Nagashima2019} (circles). The dotted lines are the fits to the experimental data. The error bars for the simulation data are obtained from five independent trajectories.
(B) Bar graphs comparing the increase in $\langle{\Delta r^2_\text{A}}(t)\rangle$ ($\langle{\Delta r^2_\text{B}}(t)\rangle$) shown in panel A (Fig.~S1A) between experiment and simulation results. (C) MSDs for the A and B loci at $t=0.5\,\text{s}$ as a function of $F$. The dotted lines are a guide to the eye. 
(D) Radial distributions of the A-locus displacement at $t=10\tau_B$ (dashed) and $t=\text{1,000}\tau_B$ (solid), compared between $F=0$ (gray) and $F=80$ (dark-red). The inset shows the 2-D projection of the trajectory of an active locus for $10^4\tau_B$ at $F=0$ and $F=80$.
}
\label{fig:msd}
\end{figure}

We calculated the MSDs separately for euchromatin and heterochromatin loci at lag time $t$, 
\begin{equation}
\langle{\Delta r^2_\mu}(t)\rangle = \frac{1}{N_\mu}\left<\sum_{i=1}^{N}\delta_{\nu(i)\mu}|\rp_i(t) - \rp_i(0)|^2\right>~,
\label{eq:msd}
\end{equation}
where $\langle \cdots \rangle$ is the ensemble average, $\nu(i)$ is the type of the $i^\text{th}$ locus, and $\delta_{\nu(i)\mu}$ is the Kronecker delta ($\delta_{xy} = 1$ if $x = y$, or 0 else) that picks out the loci of specific type $\mu$ (either A or B). 
Fig.~\ref{fig:msd}A shows that $\langle{\Delta r^2_\text{A}}(t)\rangle$ is smaller at $F=80$ compared to $F=0$. 
This result is qualitatively similar to the nucleosome MSDs measured from the interior section of the cell nucleus treated with the transcription inhibitor $\alpha$-AM \cite{Nagashima2019}. 
$\langle{\Delta r^2_\text{B}}(t)\rangle$ at $F=80$ is also smaller than at $F=0$, but the difference is marginal compared to $\langle{\Delta r^2_\text{A}}(t)\rangle$, which is consistent with the experimental results for the nuclear periphery (Fig.~S1A). 
In Fig.~\ref{fig:msd}A, the magnitude of $\langle{\Delta r^2_\text{A}}(t)\rangle$ is smaller in simulations than in experiment, where the difference  in magnitude depends  on the conversion of simulation length/time to real units (see Methods for the details). 
The level of agreement with experiments is striking, considering that we used a minimal model with no parameters to describe a very complicated phenomenon.
We could probably have obtained better agreement with experiments by tweaking the parameters in the model, which we think is not necessary for the purposes of this study. 

In Fig.~\ref{fig:msd}A, each MSD plot as a function of time is shown with  scaling behavior, $\langle{\Delta r_\text{A}^2}(t)\rangle \sim t^\alpha$. 
The scaling exponent, $\alpha$, provides the link between chromatin structure and the mobility.  
It was established previously \cite{Liu2018} that, $\alpha = \frac{2 \nu}{(2 \nu +1)}$ where $\nu$, the Flory exponent related to the radius of gyration ($R_g \sim N^{\nu}$) of the polymer chain. 
For an ideal Rouse polymer, $\nu =0.5$, and for a  polymer in a good solvent, which takes the excluded volume between the monomers into account, $\nu \approx 0.6$ \cite{Doi_Edwards1988}. 
The fit to the experimental data (gray dotted line in Fig.~\ref{fig:msd}A) when transcription is inhibited yields $\alpha \approx 0.57$, which would be consistent with the prediction for the SAW behavior. In all other cases, $\alpha$ is less than 0.5 (the value for the Rouse polymer) both in experiments and simulations, which is indicative of restricted motility.
The scaling exponent in the experimental MSD increases from $\alpha = 0.48 \,(\pm 0.05)$ to $\alpha =0.57 \,(\pm 0.07)$ upon inhibiting transcription. We estimated the errors using 10,000 samples reconstructed from the normal distribution for each data point, following the procedure described elsewhere \cite{Backlund2015}.
Our simulations capture the change in $\alpha$ such that $\Delta \alpha(\text{active}\rightarrow\text{passive}) = 0.05$ (versus 0.09 from the experiments). 
In contrast, recent experiments for chromatin dynamics in different contexts, such as loop formation and response to mechanical perturbation, reported that the MSD exponents are similar to the Rouse \cite{Gabriele2022,Keizer2022} or SAW \cite{Mach2022} polymer, \textit{i.e.}, $\alpha \gtrsim 0.5$. 
These results imply fundamentally different physical nature in transcriptionally active regions, which constrains the polymer dynamics to $\alpha < 0.5$.

In Fig.~\ref{fig:msd}B, we compare the transcription-inhibited increase ($\Delta \overline{\mathrm{MSD}}$) in the MSD, between experiment and simulations (see Eqs.~S1--S2 in SI Appendix).
We use $F=80$, which shows the smallest MSD (see Fig.~\ref{fig:msd}C), as the control.
The value of $f_0$ for $F=80$ is in the range, $f_0 \approx 3$--16 pN (see Methods), which accords well with the typical forces exerted by RNAP \cite{Yin1995}.  
Comparison between $\Delta \overline{\mathrm{MSD}}$ for the A loci (simulation) and the interior measurements (experiment) is less precise than between the B loci and the periphery.
The difference may arise because the interior measurements could include the heterochromatin contribution to some extent, whereas the periphery measurements exclude the euchromatin. 
Nevertheless, $\Delta \overline{\mathrm{MSD}}$ for all the loci is in better agreement with experiment  (Fig.~S1C).
The reasonable agreement between simulations and experiment is surprising because it is obtained {\it without adjusting any parameter to fit the data}. 
Although comparisons in Fig.~\ref{fig:msd}B are made with $F=80$, we obtain qualitatively similar results for $F$ in the range, $60 \le F \le 90$ (Fig.~S2).

The simulated MSD, at a given lag time, changes non-monotonically as $F$ changes. Remarkably, the change is confined to the A loci (Figs.~\ref{fig:msd}C and S1D--S1E); 
$\langle{\Delta r_\text{A}^2}(0.5\,\text{s})\rangle$ increases modestly as $F$ increases from zero to $F \lesssim 30$, and decreases when $F$ exceeds thirty. There is an abrupt reduction at $F \approx 50$. 
In the range, $50 \lesssim F \lesssim 80$, $\langle{\Delta r_\text{A}^2}(0.5\,\text{s})\rangle$ continues to decrease before an increase at higher $F$ values. 
In contrast, the MSD of the B loci does not change significantly with $F$ (Fig.~\ref{fig:msd}C). 
The non-monotonic trend in $\langle{\Delta r_\text{A}^2}(0.5\,\text{s})\rangle$ upon changing $F$ is reminiscent of a \textit{re-entrant} phase behavior in which a system transitions from an initial phase to a different phase and back to the initial phase as a parameter ($F$ in our case) is varied. Such a behavior occurs in a broad range of different systems \cite{Stillinger1976,Eckert2002,Banerjee2017,Henninger2021}. We present additional analyses in the next section to confirm the \textit{dynamic re-entrance}.

To obtain microscopic insights into the simulated MSDs, we calculated the van Hove function, which is simply the distribution of locus displacement at a given lag time,
\begin{equation}
G_s^\mu(\rp,t) = \frac{1}{N_\mu}\left< \sum_{i=1}^N \delta_{\nu(i)\mu} \delta (\rp + \rp_i(0) - \rp_i(t)) \right>~.
\label{eq:vanhove}
\end{equation}
Figure \ref{fig:msd}D compares the radial distribution for the displacement of A-type loci, $4\pi r^2G_s^\text{A}(r,t)$, between $F=0$ and $F=80$ at $t=10\tau_B \approx 0.007\,\text{s}$ and 1,000$\tau_B \approx 0.7\,\text{s}$ (see Methods for the unit conversion).
At short lag times, both the distributions for $F=0$ and $F=80$ show a peak at $r < 0.5\sigma$, which means that a given locus is unlikely to diffuse far from the neighboring loci in a condensed microenvironment.
At long lag times, the distribution for $F=0$ becomes more populated at $r \ge \sigma$, as the loci escape from the regime caged by the neighbors. 
The remarkably broad and flat curve for $r^2G_s^\text{A}(r,t=10^3\tau_B)$ at $F=0$, resembling a uniform distribution, signifies the heterogeneous dynamics of the A-type loci, which differs from the Gaussian distribution expected for the Rouse polymer.
The A-type loci at $F=80$ do not diffuse as much at $F=0$ even at long lag times (Fig.~\ref{fig:msd}D), and their displacements are largely within the length scale of $\sigma$. 
In contrast, there is no significant difference in $G_s^\text{B}(r,t)$ between $F=0$ and $F=80$ (Fig.~S1F).
Notably, the second peak of $G_s^\text{A}(r,t)$ at $F=80$ hints at the solid-like lattice \cite{Kim2013}.

\subsection*{Dynamic re-entrant behavior}

\begin{figure}[h!]
\centering
\includegraphics[width = 3.4 in]{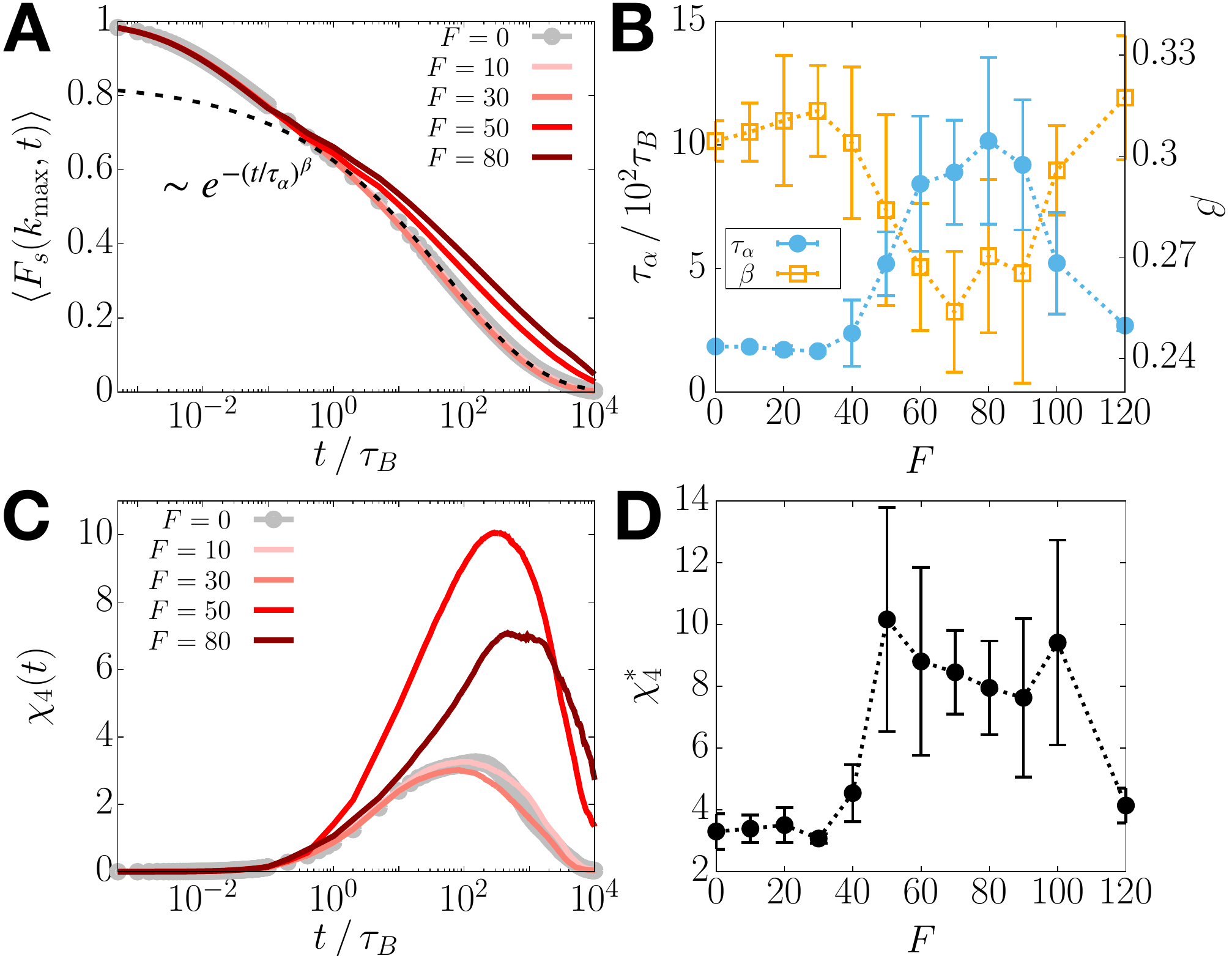}
\caption{Re-entrance in the relaxation time and dynamic heterogeneity is observed upon increasing the activity.
(A) Plot of $\langle F_s (k_\text{max},t) \rangle$ (Eq.~\ref{eq:fskt}) for different $F$. The dashed line is a stretched exponential fit for $F=0$. 
(B) $\tau_\alpha$ (blue) and $\beta$ ({orange}) of $\langle F_s (k_\text{max},t) \rangle$ as a function of $F$. The dotted lines are a guide to the eye. 
(C) Plot of $\chi_4(t)$ (Eq.~\ref{eq:chi4}) for different $F$. 
(D) The maximum value of $\chi_4(t)$ as a function of $F$.}
\label{fig:fskt}
\end{figure}

Here, we further characterize the dynamics of the CCM polymer chain in the presence of  RNAPII-induced active forces. 
The purpose is to investigate how the re-entrant behavior emerges in the CCM polymer whose dynamics is glass-like in the absence of the activity \cite{Shi2018a}.
To probe how the glass-like behavior of the polymer chain \cite{Kang2015b,Shi2018a,Liu2018} is affected by the active forces, we calculated the self-intermediate scattering function,
\begin{equation}
F_s (|\mathbf{k}|,t) = \frac{1}{N} \sum_{j=1}^N e^{i \mathbf{k}\cdot[\rp_j(t) - \rp_j(0)]}~,
\label{eq:fskt}
\end{equation}
where $\mathbf{k}$ is the wave vector. 
The scattering function, the Fourier transform of the van Hove function (Eq.~\ref{eq:vanhove}), quantifies the extent of the loci displacements over time $t$ relative to the length scale set by the wave vector ($\sim 2\pi/|\mathbf{k}|$).
We computed the ensemble-averaged $\langle F_s (k_\text{max},t) \rangle$ with $k_\text{max} = 2\pi/r_\text{LJ}^*\approx 5.60\sigma^{-1}$, where $r_\text{LJ}^*=2^{1/6}\sigma\approx1.12\sigma$ is the distance at the minimum of the non-bonding (LJ) potential (see Methods), which the nearest neighbors are likely to be located at (see the radial distribution function in Fig.~\ref{fig:struc_main} below).
The decay of $\langle F_s (k_\text{max},t) \rangle$  indicates the structural relaxation on the length scale $\gtrsim \sigma$.

Time-dependent variations in $\langle F_s (k_\text{max},t) \rangle$ (Fig.~\ref{fig:fskt}A) show stretched exponential behavior ($e^{-(t/\tau_a)^\beta}$; $\beta < 1/3$ at all $F$ values), which is one signature of glass-like dynamics.  
Note that the decay is even slower if $F$ is increased, which is in agreement with the results shown in Figs.~\ref{fig:msd} and S1.
The relaxation time, $\tau_\alpha$, calculated using $\langle F_s (k_\mathrm{max},\tau_\alpha) \rangle = 0.2$, shows that the relaxation is slowest at $F\approx80$ (Fig.~\ref{fig:fskt}B), which occurs after the dynamical transition in $\langle{\Delta r_\text{A}^2}(0.5\,\text{s})\rangle$ at $F \approx 50$ and before $\langle{\Delta r_\text{A}^2}(0.5\,\text{s})\rangle$ increases beyond $F=100$ (Fig.~\ref{fig:msd}C). 
Similarly, when the tails of $\langle F_s (k_\text{max},t) \rangle$ were fit with $e^{-(t/\tau_a)^\beta}$, the exponent $\beta$ also exhibits the analogous trend (Fig.~\ref{fig:fskt}B); that is, as $\tau_{\alpha}$ increases, $\beta$ decreases, indicating the enhancement in the extent of the glass-like behavior. 

Dynamic heterogeneity, another hallmark of glass-like dynamics \cite{Berthier2011,Kirkpatrick15RMP}, was calculated using the fourth-order susceptibility \cite{Kirkpatrick1988}, which is defined as the mean squared fluctuations in the scattering function,
\begin{equation}
\chi_4(t) = N\left[\left<F_s (k_\text{max},t)^2 \right> - \left< F_s (k_\text{max},t) \right> ^2 \right]~.
\label{eq:chi4}
\end{equation}
We find that $\chi_4(t)$, at all $F$ values, has a broad peak spanning a wide range of times, reflecting the heterogeneous motion of the loci (Fig.~\ref{fig:fskt}C). 
The peak height, $\chi_4^*$, increases till $F \approx 50$ and subsequently decreases (Fig.~\ref{fig:fskt}D). When $F$ exceeds 100, $\chi_4^*$ decreases precipitously. 
Our results suggest that there are two transitions: one at $F \approx 50$ where the dynamics slows down and the other, which is a reentrant transition beyond $F=100$, signaled by an enhancement in the mobilities of the A-type loci. 
Although the system is finite, these transitions are discernible. 

Like the MSD, when $\langle F_s (k_\mathrm{max},t) \rangle$ and $\chi_4(t)$ was decomposed into the contributions from A and B loci (Eqs.~S3 and S4), we find that the decrease in the dynamics and the enhanced heterogeneity are  driven by the active loci (Fig.~S3).
These observations, including the non-monotonicity in $\tau_\alpha$ and $\beta$ that exhibit a dynamic reentrant behavior, prompted us to examine if the dynamical changes in the A-type loci  are accompanied by any structural alterations. 

\subsection*{Transient disorder-to-order transition induced by active forces}

\begin{figure}[h!]
\centering
\includegraphics[width = 3.4 in]{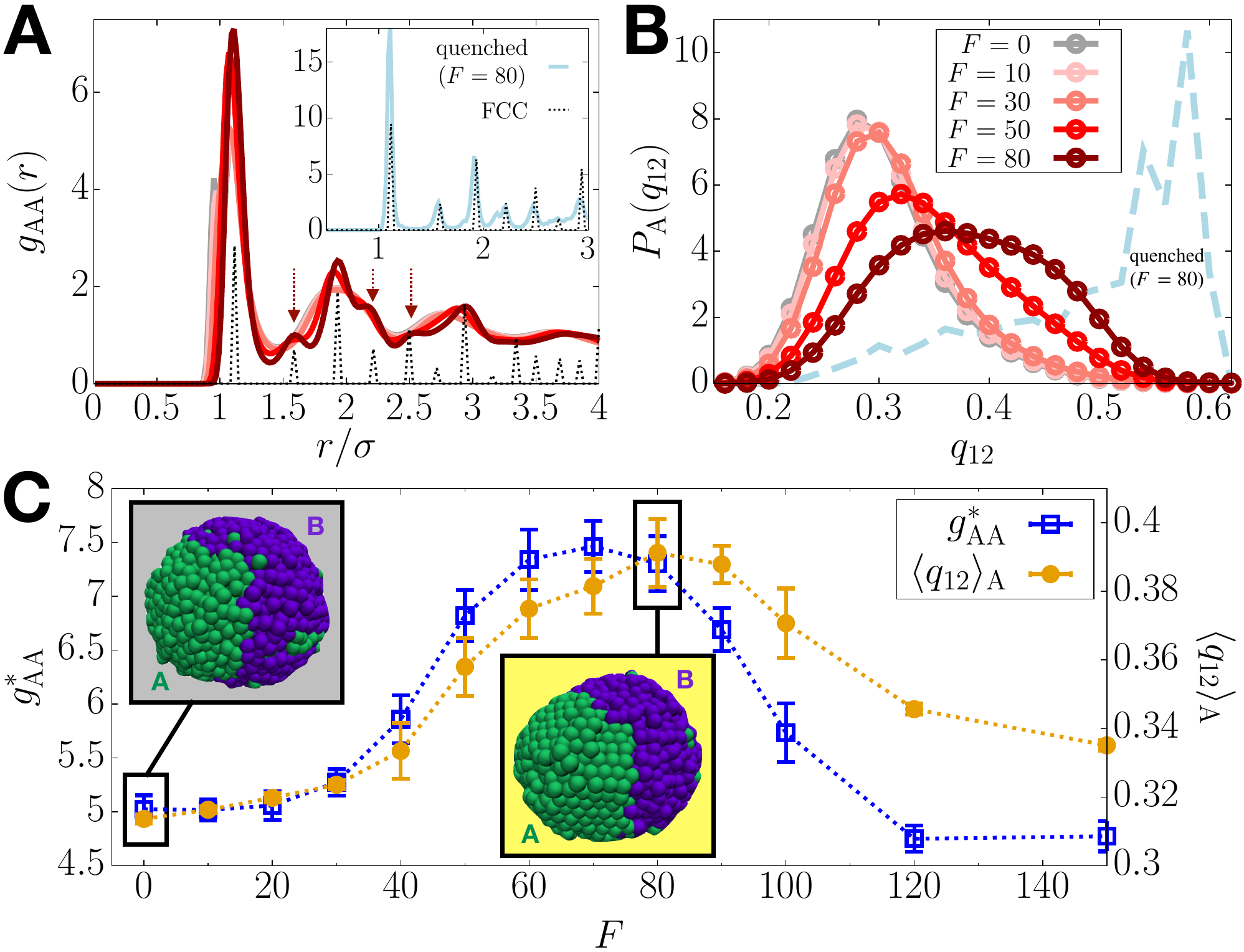}
\caption{
$F$-induced structural transition. (A) RDF for A-A pairs at different $F$ (solid lines; see the legend in panel B), where $g(r)$ for a FCC crystal is shown with the dotted line (scaled arbitrarily). The inset shows $g_\text{AA}(r)$ for the quenched polymer with $F=80$.
(B) Distributions of the BOO parameter, $q_{12}$, for A loci as a function of $F$. The dashed line is for the quenched A loci at $F=80$.
(C) Height of the dominant peak in $g_\text{AA}(r)$ (blue) and $\langle q_{12}\rangle_\text{A}$ (orange) as a function of $F$. Simulation snapshots for $F=0$ and $F=80$ are  in the gray and yellow boxes, respectively.
}
\label{fig:struc_main}
\end{figure}

The radial distribution function (RDF) for A-A locus pairs, $g_\text{AA}(r)$, with signature of a dense fluid, shows no visible change for $F \lesssim 30$ (Fig.~\ref{fig:struc_main}A). In contrast,  the height of the primary peak, $g_\text{AA}^*$, increases sharply beyond $F=30$ (Fig.~\ref{fig:struc_main}C). 
Remarkably, $g_\text{AA}(r)$ for $F=80$ exhibits secondary peaks that are characteristics of the FCC (face-centered cubic) crystal phase (Fig.~\ref{fig:struc_main}A, arrows). 
Upon further increase in $F$, these peaks disappear (Fig.~S4A) and $g_\text{AA}^*$ reverts to the level of the passive case (Fig.~\ref{fig:struc_main}C). 
In other words, the active forces, which preserve fluid-like disorder in the A loci at low $F$ values, induce a structural transition to solid-like order in the intermediate range of $F$ values, which is followed by reentrance to a fluid-like behavior at higher $F$ values.
In contrast, $g_\text{BB}(r)$ exhibits dense fluid-like behavior at all $F$ values (Fig.~S4B).
We confirm that the FCC lattice is the minimum energy configuration by determining the \textit{inherent structure} \cite{Stillinger1984, Thirumalai1989} for the A loci at $F=80$ by quenching the active polymer to a low temperature (see Methods) (Fig.~\ref{fig:struc_main}A, inset). 
In Fig.~S4C, $g_\text{AA}(r)$ at $F=0$ reflects the structure of a dense fluid upon the quench whereas $g_\text{AA}(r)$ at $F=80$ is quantitatively close to that for a perfect FCC crystal. 
Quenching does not alter the structure of the B loci at $F=80$ or $g_\text{AA}(r)$ at $F=0$ (Figs.~S4C-S4D).

To assess local order, we calculated the bond-orientational order (BOO) parameter for 12-fold rotational symmetry, $q_{12}$ (Eqs.~\ref{eq:bo_1}--\ref{eq:bo_2}) \cite{Steinhardt1983, tenWolde1995}. 
For a perfect FCC crystal, $q_{12}\approx0.6$ \cite{Omar2021}.
The distribution for A loci, $P_\text{A}(q_{12})$ (Eq.~\ref{eq:q_dist}), is centered around $q_{12} = 0.3$ at $F=0$ (Fig.~\ref{fig:struc_main}B), representing a disordered liquid state (Fig.~\ref{fig:struc_main}C, gray box).
As $F$ increases, the distribution shifts towards the right, especially in the $50 \le F \le 80$ range.
The increase of $\langle q_{12}\rangle_\text{A}$ (Eq.~\ref{eq:q_mean}) indicates a transition to a FCC-like ordered state that is visible in the simulations (Fig.~\ref{fig:struc_main}C, yellow box).
$P_\text{A}(q_{12})$ at $F=80$ is broad, whereas the inherent structure gives a narrower distribution peaked near $q_{12}=0.6$ (Fig.~\ref{fig:struc_main}B, dashed line), which shows that the ordered solid-like state coexists with the disordered fluid-like state within thermal fluctuations.
The maximum in $P_\text{A}(q_{12})$ shifts to the left for $F > 80$ (Fig.~S4E) and $\langle q_{12}\rangle_\text{A}$ decreases, suggestive of $F$-induced reentrant transition. The distribution $P_\text{B}(q_{12})$ for the B-type loci is independent of $F$ (Fig.~S4F).  
These results show that FCC-like ordering emerges in $50 \lesssim F \lesssim 100$ range. 
Outside this range, the RDFs display the characteristics of a dense fluid.
In addition, the $F$-dependent bending angle distributions for three consecutive A loci (Eq.~S5) reflect the FCC-like ordering (Fig.~S5).
The transitions in the A-type loci may be summarized as fluid $\rightarrow$ FCC/fluid $\rightarrow$ fluid, as $F$ changes from 0 to 120.

Since the overall RDF at $F=80$ does not show the FCC peaks (Fig.~S6A), the $F$-induced order may not be seen directly in the conventional experiments. 
Moreover, the ordering is transient, as implied in the coexistence of order and disorder. By transient we mean that solid-like order persists only when the polymerase induces active forces during transcription. 
The time correlation function of the BOO parameter, $C_{q_{12}}^\text{A}(t)$ (Eq.~S6), decays rapidly, at most on the order, $\approx 0.7\,\text{s}$ (Fig.~S6B), much shorter than the transcription time scale ($\sim$ minute).
More details and the visualization of the transient ordering are given in SI Appendix, Sec.~5 and Movie S1. 
Remarkably, the transient ordering preserves the large-scale structure of chromosome (see SI Appendix, Sec.~6; Fig.~S7), in agreement with the chromosome conformation capture experiments with transcription inhibition \cite{Hsieh2020,Jiang2020}. 
Here we emphasize that the solid-like nature coexisting with the liquid is sufficient to explain the experimentally observed suppression of euchromatin motions.
The predicted transient ordering could be detected by future high resolution imaging experiments. 

\subsection*{Origin of $F$-induced order}

The emergence of solid-like order in the  A-type loci is explained using the effective A-A interaction generated by $F$. 
We calculated the \textit{effective pair potential} for an A-A bond, 
\begin{equation}
u_\text{b}^\text{eff}(r) = u_\text{b}^0(r) - f_0 (r - b_0)~,
\label{eq:ub_eff}
\end{equation}
where $u_\text{b}^0(r)$ and $b_0$ are the $F$-independent bonding potential, and the corresponding equilibrium bond length, respectively. 
The $f_0 (r - b_0)$ term is the work done by the active force to stretch the bond from $b_0$. 
The equation of motion for $F \ne 0$ represents dynamics in the \textit{effective equilibrium} under the potential energy involving $u_\text{b}^\text{eff}(r)$ (Eqs.~\ref{eq:U_eff}--\ref{eq:bd_eff}).
Such effective equilibrium concept was useful for characterizing various physical quantities in active systems \cite{Loi2008,Grosberg2015,Mandal2017a,Smrek2020,Jiang2022,Omar2023}.

\begin{figure}[h!]
\centering
\includegraphics[width = 3.4 in]{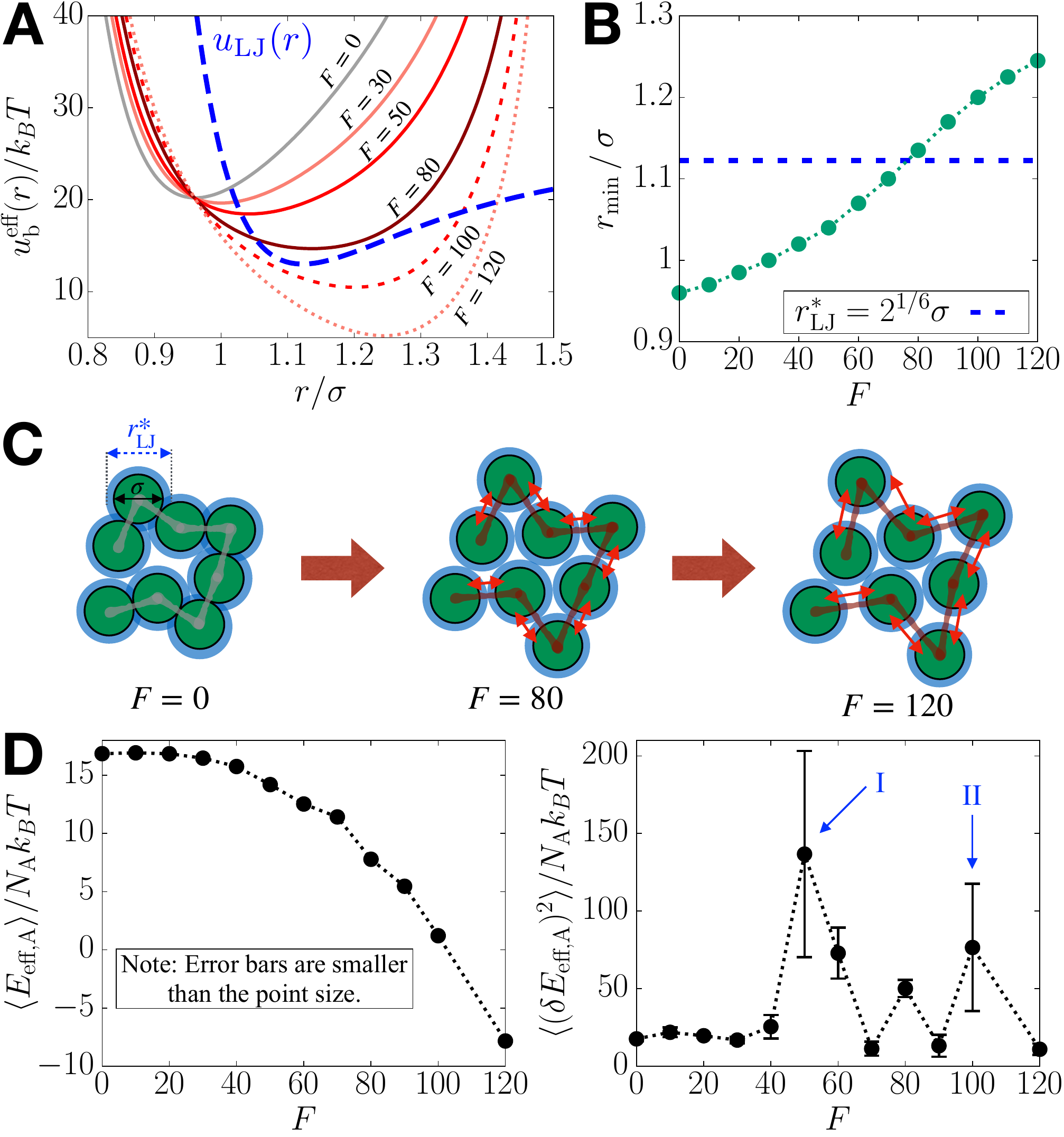}
\caption{
Effective potential energy accounts for the structural transition of the A-type loci. (A) Effective pair potential of a single A-A bond, $u_\text{b}^\text{eff}(r)$ (Eq.~\ref{eq:ub_eff}), as a function of $F$. The LJ potential, $u_\text{LJ}(r)$, which is vertically shifted, is shown in blue dashed line for comparison. 
(B) Distance at the minimum of $u_\text{b}^\text{eff}(r)$ as a function of $F$, where the dashed line indicates $r_\text{LJ}^*$ (see the main text).
(C) Schematic illustration for how the $F$-induced bond extension enables FCC ordering in the A-type loci (green circles). The blue shade shows $r_\text{LJ}^*$ at which non-bonded pairs have the most favorable interactions. The double red arrows indicate the bond extension upon increasing $F$. 
(D) Mean of the effective potential energy of A loci (left) and mean fluctuations (right) with respect to $F$. The arrows indicate the two structural transitions.
}
\label{fig:u_eff}
\end{figure}

Plots of $u_\text{b}^\text{eff}(r)$ in Fig.~\ref{fig:u_eff}A show that the effective equilibrium bond length, $r_\text{min}$, increases as $F$ increases. 
This prediction is confirmed by the direct measurement of A-A bond distance from the simulations (Figs.~S8A--S8B). 
Note that $r_\text{min}(F=80) \approx r_\text{LJ}^* \approx1.12\sigma$, where $r_\text{LJ}^*$ is the distance at the minimum of the LJ potential (Fig.~\ref{fig:u_eff}B). 
The $F$-induced extension of A-A bonds makes $r_\text{min}$ commensurate with $r_\text{LJ}^*$, which is conducive to FCC ordering \cite{Cho2021} in the active loci. 
At $F=0$, since the bond distance ($b_0 = 0.96\sigma$; see Methods) is smaller than the preferred non-bonding interaction distance, $r_\text{LJ}^*$, the lattice-matching configurations cannot be formed given the chain connectivity (Fig.~\ref{fig:u_eff}C, left).
In contrast, at $F=80$, the condition that $r_\text{min} \approx r_\text{LJ}^*$ allows the A loci to be arranged into the FCC lattice, leading to the transition to the ordered state (Fig.~\ref{fig:u_eff}C, center). 
Upon increasing the activity further (\textit{e.g.}, $F=120$), the bond distance becomes too large to be fit with the lattice structure, so the loci re-enter the disordered fluid state (Fig.~\ref{fig:u_eff}C, right). 

We can also describe the ordering behavior using thermodynamic properties based on $u_\text{b}^\text{eff}(r)$. 
We calculated the mean and variance of the effective potential energy of the A loci, $E_\text{eff,A}$ (Eq.~\ref{eq:Eeff_A}).
Fig.~\ref{fig:u_eff}D shows that $\langle E_\text{eff,A}\rangle$ decreases smoothly as $F$ changes, without pronounced change in the slope, as might be expected for a structural transition \cite{Cho2021}. 
Nevertheless, $\langle (\delta E_\text{eff,A})^2\rangle$ indicates signatures of a transition more dramatically, with  peaks at $F=50$ and $F=100$ (Fig.~\ref{fig:u_eff}D, arrows I and II). 
Thus, both ordering and reentrant fluid behavior coincide with the boundaries of the dynamic transitions noted in Figs.~\ref{fig:msd}B, ~\ref{fig:fskt}B, and \ref{fig:fskt}D.

\section*{Discussion}

We introduced a minimal active copolymer model for Chr5 in order to explain the non-intuitive experimental observation that during transcription the mobilities of the loci are suppressed. Despite the simplicity of the polymer model, we reproduce semi-quantitatively the experimental observations. 
In particular, the MSD exponent $\alpha$ ($\langle{\Delta r_\mu^2}(t)\rangle \sim t^\alpha$) decreases by 0.05 in the simulations whereas the decrease is 0.09 between the transcription inhibited and active states in the experiments (Fig. \ref{fig:msd}A). 
We find this result and the results in Fig. \ref{fig:msd} surprising because no parameter in the very simple model was adjusted to obtain the semi-quantitative agreement with experiments \cite{Nagashima2019}.   

We hasten to add that the model indeed is oversimplified, and is the major limitation of our study. The actual machinery of transcription active (or inhibited) state is extremely complicated involving the interplay of transcription factors, RNAPII, chromatin, and other cofactors.  
These elements are believed to drive the formation of a hub of connected modules resulting in condensates \cite{Hnisz2017,Cho2018,Chong2018,Sabari2018}. 
Although the structural features of chromatin in the condensates are unknown, our simulations suggest that for a very short period ($\approx$ 1s) chromatin could adopt solid-like properties when RNAPII and other transcription-associated machinery exert an active force ($\approx$ 5--10 pN) during transcription. 

\subsection*{Robustness of the conclusions} 
\begin{figure}[h!]
\centering
\includegraphics[width = 3.4 in]{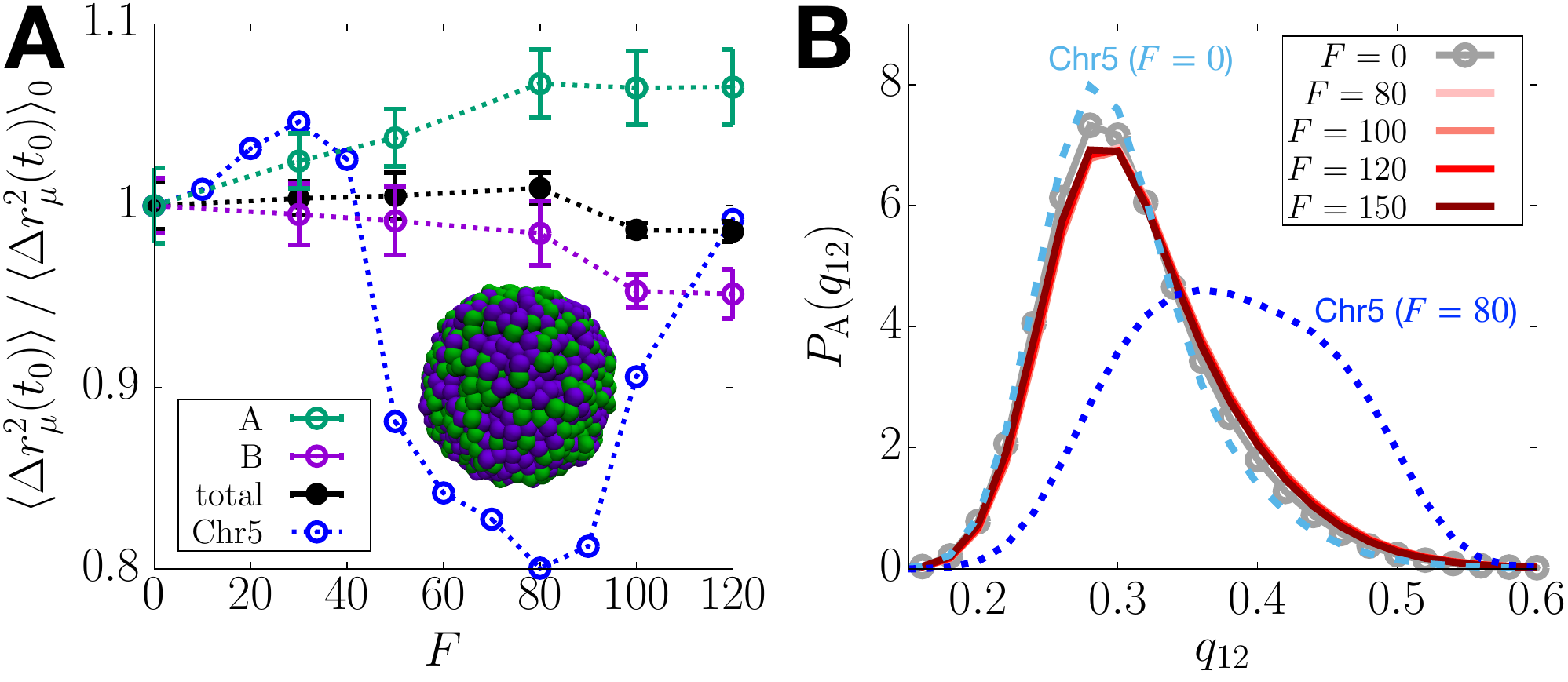}
\caption{A copolymer chain with the same length and the fraction of active loci as Chr5 but randomly shuffled epigenetic profile does not capture the experimental observations.
(A) Ratio of the MSD at $t_0=100\tau_B$ for the active random copolymer to the passive case as a function of $F$, shown in black color. The data for Chr5 is shown in blue color. The same quantities computed for the A and B loci of the random copolymer are also plotted in green and purple color, respectively. The dotted lines are a guide to the eye. 
The simulation snapshot shows that the copolymer chain is in a mixed state without discernible microphase separation.
(B) Probability distributions of $q_{12}$ for the A loci at different activity levels, where the dashed and dotted lines are the distributions for Chr5 with $F=0$ and $F=80$, respectively.
}
\label{fig:rand_seq_main}
\end{figure}

We performed three tests in order to ensure that the results are robust. 
(1) Simulations of a segment of chromosome 10, with a larger fraction of active loci, show qualitatively similar behavior (See SI Appendix, Sec.~7; Fig.~S9).
(2) For a copolymer chain whose A/B sequence is random (See SI Appendix, Sec.~8; Fig.~S10), we find that at all $F$, MSD for the A-type loci increases monotonically (Fig.~\ref{fig:rand_seq_main}A) in contrast to the non-monotonic behavior found in Chr5 with a given epigenetic (A/B) profile (Fig.~\ref{fig:msd}C). The B-type MSD decreases at high $F$ (Fig.~\ref{fig:rand_seq_main}A), which is also different from the behavior in Chr5. In addition, in the random sequences the loci exhibit fluid-like behavior at all values of $F$, and hence cannot account for the experimental observation (Fig.~\ref{fig:rand_seq_main}B). 
Thus, $F$-induced decrease in the motility of the A loci, accompanied by transient ordering, occurs only in  copolymer chains exhibiting microphase separation between A and B loci---an intrinsic property of interphase chromosomes \cite{Nagano2017}. In other words, the epigenetic profile determines the motilities of the individual loci (or nucleosomes) during transcription.
(3) We performed simulations by removing all the loop anchors (analog of topological constraints \cite{Smrek2020}; see SI Appendix, Sec.~9). 
Fig.~S11 shows that the $F$-induced dynamical slowdown and ordering is not significantly affected by the loop anchors. 
The modest increase in the MSD upon removal of loops, at $F=0$, is qualitatively consistent with the experimental measurements of increased chromatin mobility when cohesin loading factors were deleted \cite{Nozaki2017}. 
The experimental results suggest that a reduction in transcriptional activity may be induced by loss of loops, which is assumed to be a consequence of depletion of cohesin (see ``Connection to other experiments'' below for more discussion).
These tests show that our conclusions are robust, and the model might serve as a minimal description of the effect of transcription-induced active forces on real chromosomes.

\subsection*{Emergence of solid-like ordering induced by active forces} 
In the single copolymer chain, which reflects the compact nature of interphase chromosomes \cite{Manuelidis1990,Lieberman-aiden2009}, transient solid-like ordering would occur only if the distance between two bonded loci is roughly equal to the distance at which the pair potential between the non-bonded loci is a minimum (for the LJ potential, $r_\text{LJ}^* = 2^{1/6}\sigma$) \cite{Cho2021}. 
This argument is generic because it only depends on the presence of attractive interactions between non-bonded loci, which is needed to ensure that interphase chromosomes are compact.
Application of the active forces increases $r_\text{min}$, and over a small range of forces $r_\text{min}$ becomes commensurate with $r_\text{LJ}^*$ (Figs.~\ref{fig:u_eff}A--\ref{fig:u_eff}C). 
It should be noted that the criterion for solid-like ordering is {\it independent} of the interaction parameters characterizing the chromosome, and depends only on the magnitude of the force (see SI Appendix, Sec.~10; Fig.~S12).
If the bonds are rigid, chain segments in the polymer chain, under confinement and active extensile forces, would align exhibiting a nematic order \cite{Saintillan2018}. 
Thus, the mechanisms of our FCC-like ordering and the nematic order in Ref.~\citenum{Saintillan2018} are qualitatively different. However, both the studies show that  active forces induce ordering in chromatin. Additional comparisons between the two models are given in SI Appendix, Sec.~11. 
 
\subsection*{Studies of activity-induced changes in polymers} 
The roles of active processes and associated forces in chromosome organization and dynamics have been probed in theoretical studies based on polymer models \cite{Ganai2014,Liu2018,Saintillan2018,Liu2021,Mahajan2022a,Jiang2022,Goychuk2023}.
It was shown that athermal random noise, when applied selectively on the active loci, affects the nuclear positioning of the active loci \cite{Ganai2014,Liu2018,Jiang2022}.  
Interestingly, using a combination of analytical theory and simulations \cite{Goychuk2023}, it has been shown that a modest difference in activity between active and inactive loci leads to microphase separation \cite{Goychuk2023}, a key feature of organization in interphase chromosomes.   
Using a model that is similar to that used here \cite{Jiang2022}, it has been shown that active forces are important in driving heterochromatin to the periphery of the nucleus, which attests to the dynamic nature of  chromatin structures. 
Typically applying active forces resulted in the increased MSD of the polymer chain \cite{Bianco2018,Put2019,Foglino2019,Goychuk2023}, which would not capture the suppressed chromatin mobility in human cells with active transcription \cite{Nagashima2019,Shaban2020,Zidovska2013}.  
Our minimal model, in the absence of active forces, leads to compact globule that exhibits glass-like dynamics over a long time  \cite{Shi2018a}, which may be needed to reproduce the experimental findings.  
Furthermore, it is likely that explicit inclusion of the precise epigenetic states, as done here for chromosomes 5 and 10, is required for observing the unexpected role of active forces on locus mobilities.

\begin{figure}[h!]
\centering
\includegraphics[width = 3.4 in]{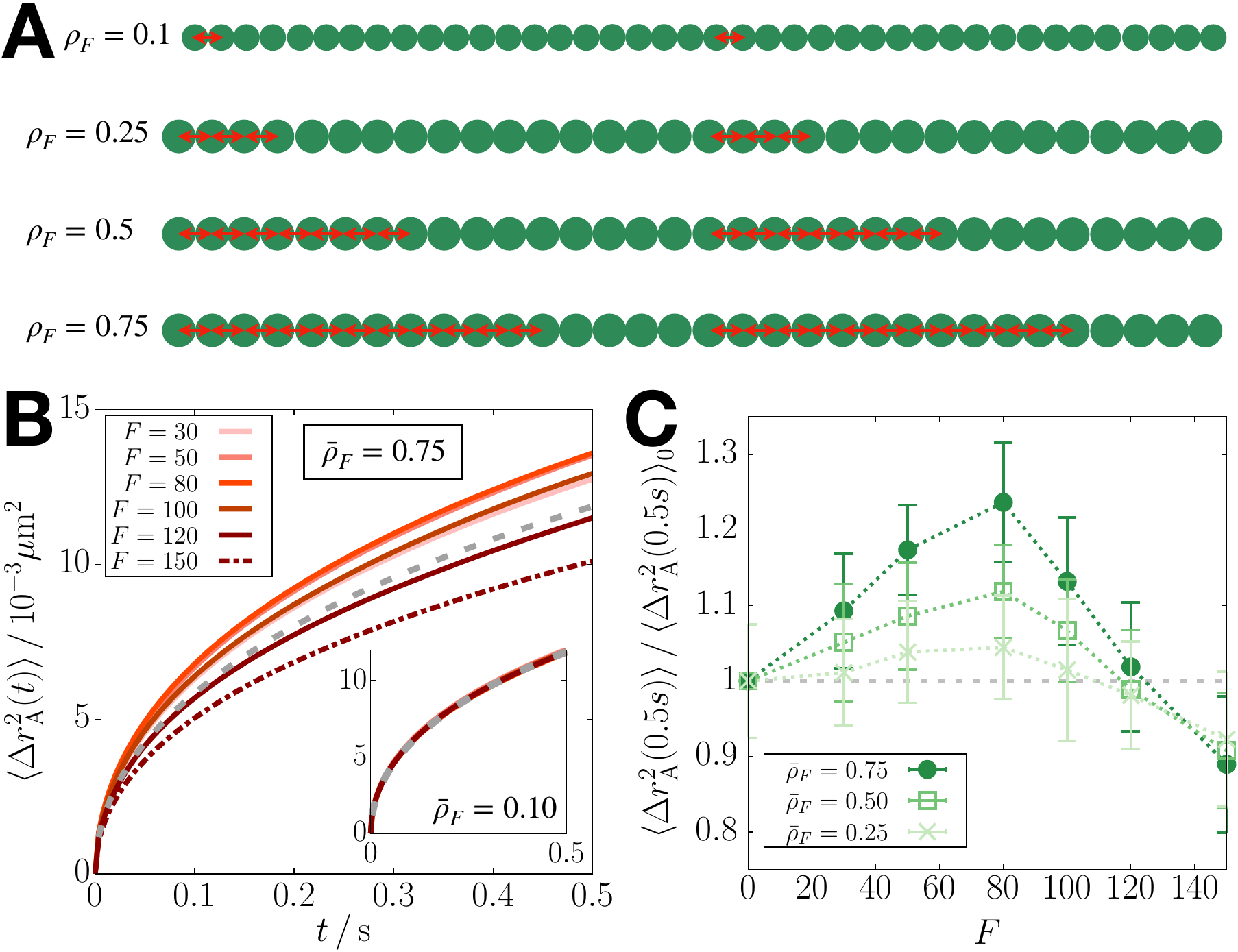}
\caption{Effect of  density of active forces on the mean square displacement.
(A) Schematic illustration  implementing the active forces at a given density in a single gene region of 40 ($\rho_F = 0.1$) or 32 ($\rho_F \ge 0.25$) loci, where the green circles are the gene loci and the red double-arrows are the active forces. 
(B) $\langle{\Delta r^2_\text{A}}(t)\rangle$ from the simulations at the average force density $\bar{\rho}_F = 0.75$ a function of $F$, where the gray dashed line is $\langle{\Delta r^2_\text{A}}(t)\rangle$ at $F=0$. The inset shows the results for $\bar{\rho}_F = 0.1$.
(C) Ratio of the MSD at 0.5 s for the active polymer at given force density to the passive case as a function of $F$. The dotted lines are a guide to the eye.
}
\label{fig:f_den}
\end{figure}

\subsection*{Effect of active force density}   
To assess how the force density affects the results, we repeated the simulations for Chr5 by adapting the model such that the active forces are applied to only a fraction of the A-type loci. 
First, we applied the extensile force dipole on a bond of A-type loci per 20 loci in each gene region (see Fig.~\ref{fig:f_den}A for schematic illustration). 
This implementation corresponds to the RNAPII density of $\sim0.04$ per kb, assuming that there is a linear relation between the force density and the linear RNAPII density. 
At this density of the active loci, there is no discernible change in the polymer dynamics regardless of the force magnitude (Fig.~\ref{fig:f_den}B, inset).

We then increased the force density gradually to detect any changes in the polymer dynamics. 
We describe the force density using $\rho_F (i)=n_F (i)/n_\text{A} (i)$, where $n_F (i)$ is the number of loci subject to active forces in the $i^\text{th}$ gene and $n_\text{A} (i)$ is the total number of loci in the gene. 
For instance, the above case in which the forces were applied on a bond every 20 loci corresponds to $\rho_F=0.1$, whereas $\rho_F=1$ for our original simulations (yielding the results shown in Figs. 2--6). 
We considered the cases with the average force density, $\bar{\rho}_F= 0.25$, 0.5, and 0.75 (Fig.~\ref{fig:f_den}A). 
At these force densities, the MSD shows a non-monotonic change upon increasing $F$, which differs from that at $\rho_F=1$ (Figs.~\ref{fig:f_den}B--\ref{fig:f_den}C). 
The MSD increases for $F\lesssim80$ and decreases for $F>80$. 
The increase is comparable to that for $\rho_F=1$ at $F\approx30$. 

Our model requires a large density of active forces ($\rho_F=1$) in order to capture the observed decrease in the chromatin loci mobility upon transcription activation \cite{Nagashima2019,Zidovska2013}.
This restriction may have two implications. 
On the one hand, this might be considered as a limitation of the model. 
Due to its simplicity, the model might not be able to account for the effect of RNAPII-induced force at a reasonable linear RNAPII density. 
However, what is more relevant may be  the number of RNAPII molecules per unit volume during transcription, which is harder to estimate. Moreover, it is unclear how much active force is transmitted by  RNAPII molecules to the gene loci at the resolution used in the model.  
On the other hand, the simulation results imply that the loci exhibiting mobility decrease upon transcription activation might experience more force than expected from active RNAPII molecules. 
More precisely, collectively increased bond distances between consecutive gene loci will result in the motility decrease due to the solid-like ordering (see Fig.~\ref{fig:u_eff}C). 
Such extra amount of force or extended bonds could arise from ATP-dependent chromatin/nucleosome remodeling that is associated with active transcription \cite{Becker2002}. 
Similarly, our model suggests that if there is less amount of force or bond extension in gene regions (\emph{i.e.}, $\rho_F\lesssim0.75$), the mobility of the gene loci will increase upon activation without  ordering, which can account for the results from Ref.~\citenum{Gu2018} (see the next subsection).

\subsection*{Connection to other experiments}
Gu \textit{et al.}~\cite{Gu2018} reported that transcription enhances the mobility of specific gene promoters/enhancers, which seems to contradict the results in a related experimental study \cite{Nagashima2019,Zidovska2013}. Gu \textit{et al.}  tracked the motion of only a few loci whereas  Nagashima \textit{et al.}~\cite{Nagashima2019} measured the motility of many loci across the nucleus. 
In addition, the active genes chosen by Gu \textit{et al.} seem to be outliers in terms of changes in the gene expression level upon  cell differentiation \cite{Gu2018}. 
Those genes show about 50--100 times difference in their expression levels upon activation or suppression following  cell differentiation, although the expression of other genes mostly change by a factor of 4--5.
We surmise that the remarkably high level of gene transcription activity results in the mobility changes that are  opposite to the global trend for all other genes \cite{Nagashima2019,Shaban2020,Zidovska2013}. 
However, the precise explanation for the differences between the experiments \cite{Gu2018,Nagashima2019,Zidovska2013} is lacking.
Using our results, we interpret, tentatively, that the enhanced motility observed for the activated genes \cite{Gu2018} likely corresponds to $0\le F \le30$ or $90\le F \le120$ at $\rho_F = 1$ (see Fig.~\ref{fig:msd}C), or $0\le F \lesssim 80$ at $\rho_F \lesssim 0.75$ (see Fig.~\ref{fig:f_den}C), or possibly to the case where these genes are situated with no other co-activated genes in their proximity (see Discussion below).  
On the other hand, a majority of the active loci tracked in Nagashima \textit{et al.} \cite{Nagashima2019} likely corresponds to $60\le F \le90$ at $\rho_F = 1$.

In another experiment \cite{Nozaki2017}, it was shown that acute depletion of cohesin, which mediates the formation of chromatin loops along with CTCFs \cite{Rao2014,Schwarzer2017,Rao2017,Kim2019}, results in  increased motility of nucleosomes. 
Our polymer simulations capture the enhanced dynamics upon  loop loss at $F=0$. 
Indeed, when we compare the MSDs between $F=80$ with loops and $F=0$ without loops, a near quantitative agreement with the experiment ($\sim$20--25\% increase) is obtained. 
This result implies that the loop loss may lead to the decreased transcription activity of some genes, possibly due to the disruption of interactions between the cognate promoter and enhancer. 
Although the downregulation of genes was observed upon  loop loss \cite{Schwarzer2017,Rao2017,Hsieh2022}, the overall effect on gene expression is not clearly understood since the loop formation is dynamic \cite{Hansen2018,Gabriele2022,Mach2022}, and is likely coupled to transcription \cite{Banigan2023,Zhang2023}. 
More precise theory that incorporates the effect of dynamic loops on transcription needs to be constructed using the experiments  that measure locus motion and real-time gene expression simultaneously \cite{Bruckner2023}.

The physical picture suggested here needs additional validation by future experiments. 
Direct observation of transient ordering could be challenging in the resolution limit of currently available in imaging methods. 
Possibly, the live-cell magnetic tweezer technique \cite{Keizer2022} may be utilized to test our finding. 
Recent experiments support our conclusions indirectly \cite{Strickfaden2020,Bignaud2022,Bohrer2023}. 
In particular, Bohrer and Larson \cite{Bohrer2023} highlighted the significant correlation between the spatial proximity and the co-expression of genes by  analyzing the DNA and RNA FISH data \cite{Su2020}.  
The role of gene co-activation is implicitly considered in  our study because we apply active forces on all the gene-transcribing loci in a given chromosome region. 
We do not observe the activity-induced decrease in chromatin mobility when only one gene or a few loci are active (see SI Appendix, Sec.~12; Fig.~S13). Based on these simulations, we conclude tentatively that motility of loci would increase \cite{Gu2018} when only a single gene is transcribed whereas RNAPII-induced forces would suppress chromatin loci mobility when a few genes are co-expressed.

\subsection*{Limitations} 
Our model is not without limitations. 
For instance, explicit effect of transcription factors may bring the theory closer to experiments \cite{Wagh2023} but will come at the expense of complicated simulations.
In addition, our model does not include hydrodynamic effects \cite{Bruinsma2014,Saintillan2018,Eshghi2022} that contribute significantly to the correlated motion of chromatin observed in the experiments \cite{Zidovska2013}. 
On the experimental side \cite{Nagashima2019}, the mobility change might arise due to DNA damage that could be induced by transcriptional inhibition \cite{Lee2002,Zidovska2013}.
The DNA repair process, which is not considered here, could be an alternate mechanism that could explain the increase in the MSD \cite{Zidovska2013,Eaton2020}. 

For a more realistic picture of gene expression dynamics, our model should be extended to incorporate  RNAP-induced force transmission in temporally and/or spatially dependent manner. 
This extension may also include the implementations of the features at higher resolution, such as the dynamic loop extrusion \cite{Nuebler2018,Gabriele2022,Mach2022,Zhang2023} and the RNAP translocation with nascent RNAs elongated \cite{Henninger2021}. 
Such a  biophysical model would provide clearer pictures on how the gene expression is regulated in conjunction with the dynamically-evolving chromatin topology.
 
The prediction of transient solid-like ordering might be criticized because currently there is no experimental validation of the proposed mechanism. However,  whether chromatin is solid-like \cite{Strickfaden2020} or liquid-like \cite{Nozaki2017} continues to be controversial \cite{Zidovska2020}. It could adopt both these characteristics depending on length and time scales \cite{Nozaki2023}. In light of the unresolved nature of the dynamical state of  chromatin, our proposal of the emergence of  transient solid-like order is not unreasonable, especially considering that the simulations confirm the non-intuitive experimental observations. Additional experiments are needed to clarify the nature of the chromatin state, which is likely to change depending on the chromatin activity. 

\matmethods{

\subsection*{Assignment of the monomer type and loop anchors in CCM}

We model chromosomes as a copolymer that embodies the epigenetic information. 
The locus type and loop anchor locations (see Fig.~\ref{fig:exp_model}) in the CCM are assigned following the procedures given in our previous study \cite{Shi2018a}.
The locus type (euchromatin or heterochromatin) is assigned using the epigenetic information from the Broad ChromHMM track \cite{Rosenbloom2013,Ernst2010,Ernst2011}.
The first 1--11 states are labeled as active (euchromatin, A) and states 12--15 as inactive (heterochromatin, B). 
To identify the loop anchors, we use the locations of unique CTCF loops identified by Rao et al. \cite{Rao2014}. 
We assign loop anchors to the loci whose genomic regions are closest to the centroids of the experimentally identified anchors.  
For the results shown above, we considered chromosome 5 (Chr5) in the region from 145.87 to 150.67 Mb from the GM12878 cell line with the resolution of 1.2 kb per locus.
The numbers of active and inactive loci are $N_\text{A} = 982$ and $N_\text{B} = 3018$, respectively. 
In this Chr5 region, there are 13 CTCF loops. 
The simulations were performed using a single copolymer model of Chr5.

\subsection*{Potential energy function of CCM}

The CCM energy function for a chromosome is given by $U_0 = U_\text{NB} + U_\text{B} + U_\text{L}$, where $U_\text{NB}$, $U_\text{B}$, and $U_\text{L}$ represent the energetic contributions of non-bonding interactions, bond stretch, and CTCF-mediated loops, respectively. We use the Lennard-Jones potential for the non-bonding pair interactions, given by $U_\text{NB} = \sum_{i=1}^{N-2}\sum_{j=i+2}^{N} u_\text{LJ}(r_{i,j} | \epsilon_{\nu(i)\nu(j)}, \sigma)$, where
\begin{equation}
u_\text{LJ}(r|\epsilon,\sigma) = 4\epsilon \left[\left(\frac{\sigma}{r}\right)^{12} - \left(\frac{\sigma}{r}\right)^6 \right]~,
\label{eq:LJ}
\end{equation}
$r_{i,j} = |\rp_i - \rp_j|$ is the distance between the $i^{th}$ and $j^{th}$ loci, and $\sigma$ is the diameter of a locus. 
The diameter of A and B type loci are the same. 
Here, $\nu(i)$ is the locus type, A or B, so there are three interaction parameters, $\epsilon_\text{AA}$, $\epsilon_\text{BB}$, and $\epsilon_\text{AB}$.
The bond stretch energy is written as $U_\text{B} = \sum_{i=1}^{N-1} u_\text{b}^0(|\bp_i|) $, where $u_\text{b}^0(r) = u_\text{\tiny FENE}(r) + u_\text{\tiny WCA}(r)$ and $\mathbf{b}_i = \mathbf{r}_{i+1}-\mathbf{r}_{i}$. 
$u_\text{\tiny FENE}(r)$ is the finite extensible nonlinear elastic (FENE) potential \cite{Warner1972,Kremer1990},
\begin{equation}
u_\text{\tiny FENE}(r) =  -\frac{1}{2} K_\text{S} r_\text{max}^2 \ln\left(1 - \frac{r^2}{r_\text{max}^2} \right)~,
\label{eq:fene}
\end{equation}
where $r_\text{max}$ is the maximum bond length and $K_\text{S}$ is the FENE spring constant. $u_\text{\tiny WCA}(r)$ is the Weeks-Chandler-Anderson (WCA) potential \cite{Weeks1971}, given by,
\begin{equation}
u_\text{\tiny WCA}(r) =\big[u_\text{nb}(r|\epsilon_\text{b},\sigma)+\epsilon_\text{b}\big]\Theta(r_\text{LJ}^* - r)~,
\label{eq:wca}
\end{equation}
where $\Theta(x)$ is the Heaviside step function ($\Theta(x) = 1$ if $x > 0$ and $\Theta(x) = 0$ if $x \le 0$). 
The WCA potential is the repulsive tail of the LJ potential, which is shifted and truncated at $r_\text{LJ}^* = 2^{1/6}\sigma$, and keeps a pair of adjacent loci from collapsing. 
Finally, the constrains imposed by  the loop anchors is modeled using a harmonic potential,
\begin{equation}
U_\text{L} = \sum_{\{p,q\}} K_\text{L} (r_{p,q} - a)^2~,
\label{eq:U_L}
\end{equation}
where $\{p,q\}$ is the set of indices of the loop anchors, $a$ is the equilibrium bond length between the pair of loop anchors, and $K_\text{L}$ is the harmonic spring constant.

The values of the parameters in $U_\text{B}$ and $U_\text{L}$ are $K_\text{S} = 30 k_B T/\sigma^{2}$, $r_\text{max} = 1.5\sigma$, $\epsilon_\text{b} = 1.0k_B T$, $K_\text{L} = 300  k_B T/\sigma^{2}$, and $a = 1.13\sigma$.
With these parameter values, the equilibrium bond length is given by $b_0 = 0.96 \sigma$ at which $u_\text{b}^0(r)$ is minimized. 
For the energetic parameters in $U_\text{NB}$, we used $\epsilon_\text{AA} = \epsilon_\text{BB} = 2.4 k_BT$ and $\epsilon_\text{AB} = \frac{9}{11}\epsilon_\text{AA}$. 
The previous study showed that the CCM with this parameter set reproduces the Hi-C inferred contact maps for chromosomes 5 and 10 well \cite{Shi2018a}.  
In particular, compartment features on $\sim5$ Mb scale and topologically associating domains (TADs) on $\sim0.5$ Mb are reproduced using the CCM simulations. 
In addition, the CCM simulations reproduced the dynamic properties that agree with experimental results.

\subsection*{Simulation details with active force}
We performed Brownian dynamics simulations by integrating the following equation of motion, which for the $i^\text{th}$ locus is given by, 
\begin{equation}
\zeta\dot{\rp}_i = -\frac{\partial}{\partial \mathbf{r}_i}U_0(r^N(t)) + \mathbf{R}_i(t) + \delta_{\nu(i)\text{A}}\mathbf{f}_i(t)~,
\label{eq:bd}
\end{equation}
where $\zeta$ is the friction coefficient and $U_0(r^N(t))$ is the potential energy of the CCM polymer chain with the configuration, $r^N(t) = \{\rp_1(t),\cdots,\rp_N(t)\})$.
The Gaussian white noise, $\mathbf{R}_i(t)$ satisfies $\langle \mathbf{R}_i(t)\cdot \mathbf{R}_j(t')\rangle = 6\zeta k_B T\delta_{ij}\delta(t-t')$, where $k_B$ is the Boltzmann constant and $T$ is temperature. 
The Kronecker delta, $\delta_{\nu(i)\text{A}}$, in the last term of Eq.~\ref{eq:bd} ensures that $\mathbf{f}_i(t)$ acts only on the A-type loci.
The exact form of $\mathbf{f}_i(t)$ is, 
\begin{equation}
\mathbf{f}_i(t) = f_0 \left[\delta_{\nu(i-1)\text{A}}\mathbf{\hat{b}}_{i-1}(t) - \delta_{\nu(i+1)\text{A}}\mathbf{\hat{b}}_i(t) \right]~. 
\label{eq:f_a}
\end{equation}

The time step for integrating Eq.~\ref{eq:bd} was chosen as $\delta t = 10^{-4} \tau_B$, where $\tau_B$ is the Brownian time, defined by $\tau_B = \sigma^2/D_0$.
Here, $D_0 = k_B T/\zeta$ is the diffusion coefficient of a single locus. 
Using the Stokes-Einstein relation, $D_0 = k_B T /6 \pi\eta R$, where $\eta$ is the viscosity of the medium and $R$ is the radius of a locus, we can evaluate the diffusion coefficient and the simulation time step in real units. 
We choose $\eta = 0.89\times10^{-3}\,\text{Pa}\cdot\text{s}$ from the viscosity of water at 25$^\text{o}$C and $R\approx \sigma/2$. 
We take $\sigma = 70$ nm from an approximate mean of the lower and upper bounds for the size of 1.2 kb of chromatin including six nucleosomes and linker DNAs, which are estimated as 20 nm and 130 nm, respectively \cite{Shi2018a}.
Hence, in real time and length units,  $D_0 = \frac{(1.38 \times 10^{-23}\,\mathrm{Pa\cdot m^3/K})(298 \mathrm{K})}{6\pi(0.89\times10^{-3}\,\text{Pa}\cdot\text{s})(35\times 10^{-9} \,\text{m})}\approx 7.0 \,\mu \mathrm{m^2 /s}$
and $\tau_B \approx 0.7$ ms.

For the most part, the results of the simulations are reported in the reduced units. 
The energy parameters are given in units of $k_B T$. Thus, the fundamental energy unit is  $\epsilon = k_B T$, which means that the reduced temperature becomes $T^* = Tk_B/\epsilon = 1$.
The fundamental length and time units are $\sigma$ and $\tau=(m\sigma^2/\epsilon)^{1/2}$, respectively, where $m$ is the mass of a single locus.
With an estimate of 260 kDa per nucleosome with 200 bps of dsDNA, we obtain $\tau \approx 55.5$ ns in real units so the time step is given by $\delta t \approx 1.26 \tau$.
The magnitude of the active force is  $F= f_0 \sigma / \epsilon= f_0 \sigma / k_B T$. 
For instance, $F=80$ corresponds to 3 to 16 pN in real units, based on the lower and upper bounds of $\sigma$ specified above (20 nm to 130 nm). 

Each simulation starts from a collapsed globule configuration that is obtained from the equilibration of an extended polymer chain using the low-friction Langevin thermostat at $T^*=1.0$ in the absence of active force. 
We propagate the collapsed configuration with active force for $10^6\delta t$ until the active polymer reaches a steady state where the radius of gyration and potential energy does not increase further. 
Subsequently, we run the simulation for additional $2 \times 10^8\delta t$ and generate five independent trajectories to obtain the statistics needed to calculate the quantities of interest. 
All the simulations were performed using LAMMPS \cite{Plimpton1995}.

\subsection*{Inherent structure for the active polymer}

Following the concept introduced by Stillinger and Weber \cite{Stillinger1984}, we investigated the \textit{inherent structure} for each locus type in the CCM. 
The inherent structure is the ideal structure in molecular liquid which is preferred by the geometrical packing of particles, so it is the arrangement of particles expected by removing thermal excitations. 
Practically, the inherent structure is determined by minimizing the energy of the system using the steepest descent method \cite{Stillinger1985}.
This procedure, however, is not directly applicable to our system because it involves the active force that is not derived from a given potential energy.
Instead, we perform the quench of the CCM with the active force to low temperature. 
In a previous study, it was shown that quenching the monodisperse colloidal liquid generates a stable BCC (body-centered cubic) crystal, which cannot be easily obtained by the standard steepest descent \cite{Thirumalai1989}.

To quench the system to a low temperature, we took the configuration from $T=298 K$ $(T^* = 1.0)$ and ran Brownian dynamics simulation at $T \approx 3\,\text{K}$ $(T^* = 0.01)$. 
The Brownian time and simulation time step are scaled accordingly such that $\tau_B' = 100\tau_B$ and $\delta t' = 10^{-4}\tau_B' = 100\delta t = 126\tau$.
The quench simulation was run for $10^6 \delta t'$ during which the system energy was minimized until it reached a plateau value within $\sim10^5 \delta t'$. 
We computed the radial distribution functions, $g_\text{AA}(r)$ and $g_\text{BB}(r)$, from the quenched configurations.

\subsection*{Bond-orientational order parameter}

Following Ref.~\citenum{Steinhardt1983}, we computed the bond-orientational order (BOO) parameter for each individual locus. 
The BOO with $l$-fold symmetry for the $i^\text{th}$ locus is defined as,
\begin{equation}
q_{l}(i)= \left[\frac{4\pi}{2l+1} \sum_{m= -l}^{l} |q_{lm}(i)|^2 \right]^{1/2},
\label{eq:bo_1}
\end{equation}
where $q_{lm}(i)$ is the average of the spherical harmonics, $Y_{lm}$, for the bond angles formed between the $i^\text{th}$ locus and its nearest neighbors,
\begin{equation}
q_{lm}(i)= \frac{\sum_{j\neq i}^{N} Y_{lm}(\rp_j - \rp_i) \Theta(r_c - r_{i,j})}{\sum_{j\neq i}^{N} \Theta(r_c - r_{i,j})}~.
\label{eq:bo_2}
\end{equation}
We use $r_c = 1.4\sigma$, where the radial distribution function has the local minimum after the peak corresponding to the first coordination shell, as the cutoff pair distance for the nearest neighbors.
The probability distribution and average of BOO parameter were computed depending on the locus type such that 
\begin{equation}
P_\mu(q_{12}) = \frac{1}{N_\mu}\sum_{i=1}^N \delta_{\nu(i)\mu}\langle\delta(q_{12}(i)-q_{12})\rangle~,
\label{eq:q_dist}
\end{equation}
and 
\begin{equation}
\langle q_{12} \rangle_\mu =  \frac{1}{N_\mu}\sum_{i=1}^N \delta_{\nu(i)\mu} \langle q_{12}(i) \rangle~.
\label{eq:q_mean}
\end{equation}
The distributions, $P_\text{A}(q_{12})$ and $P_\text{B}(q_{12})$, for different values of $F$ are shown in Figs.~S4E and S4F.

\subsection*{Effective potential energy}

Since the active force is not stochastic, it may be treated as a pseudo-conservative force, which contributes to potential energy. 
In particular, we define the effective potential energy, 
\begin{equation}
U_\mathrm{eff} = U_0 - f_0\sum_{i=1}^{N-1} \delta_{\nu(i)\text{A}}\delta_{\nu(i+1)\text{A}}(|\mathbf{b}_i|-b_0)~,
\label{eq:U_eff}
\end{equation}
where the second term, denoted by $U_a$, represents the work due to the active force on the A-A bonds. 
It can be verified that $(\partial/\partial \mathbf{r}_i) U_\mathrm{a} = - \delta_{\nu(i)\text{A}}\mathbf{f}_i$, so the equation of motion (Eq.~\ref{eq:bd}) can be rewritten as, 
\begin{equation}
\zeta\mathbf{\dot{r}}_i =   -\frac{\partial}{\partial \mathbf{r}_i} U_\mathrm{eff}(r^N(t)) + \mathbf{R}_i(t)~.
\label{eq:bd_eff}
\end{equation}
The minus sign in $U_a$ indicates that this potential prefers bond extension. 
Note that $U_a$ only affects the bond potential energy of the A-A bonds. 
It is consistent with the effective bond potential, $u_\text{b}^\text{eff}(r)$ in Eq.~\ref{eq:ub_eff}, with $U_a = \sum_{i=1}^{N-1} \delta_{\nu(i)\text{A}}\delta_{\nu(i+1)\text{A}} [u_\text{b}^\text{eff}(|\mathbf{b}_i|) - u_\text{b}^0(|\mathbf{b}_i|)]$.
In Fig.~\ref{fig:u_eff}D, we showed the mean and variance of the effective potential energy for the A loci ($E_\text{eff,A}$) as a function of $F$. 
$E_\text{eff,A}$ is given by the sum of all the pairwise interactions involving the A loci, or the total effective potential energy excluding the contributions of B-B pairs,
\begin{eqnarray}
E_\text{eff,A} = U_\text{eff} - \sum_{i=1}^{N-2}\sum_{j= i+2}^{N} \delta_{\nu(i)\text{B}}\delta_{\nu(j)\text{B}}u_\text{LJ}(r_{i,j} | \epsilon_{\nu(i)\nu(j)}, \sigma) \nonumber\\
- \sum_{i=1}^{N-1}\delta_{\nu(i)\text{B}}\delta_{\nu(i+1)\text{B}} u_\text{b}^0(|\mathbf{b}_i|)\nonumber\\
- \sum_{\{p,q\}}\delta_{\nu(p)\text{B}}\delta_{\nu(q)\text{B}} K_\text{L} (r_{p,q} - a)^2~.
\label{eq:Eeff_A}
\end{eqnarray}
}
\showmatmethods{} 

\acknow{We thank Alexandra Zidovska, Bin Zhang, Debayan Chakraborty, Xin Li, and Davin Jeong for useful discussions.  We thank the National Science Foundation (CHE19-00093) and the Welch Foundation through the Collie-Welch Chair (F-0019) for support.}

\showacknow{} 

\pagebreak
\onecolumn
\clearpage
\noindent{\bf\sffamily\LARGE Supporting Information Text} 
\large

\titleformat{\section}
  {\Large\sffamily\bfseries}
  {\thesection.}
  {0.5em}
  {#1}
  []

\setcounter{equation}{0}
\setcounter{figure}{0}
\setcounter{table}{0}
\renewcommand{\theequation}{S\arabic{equation}}
\renewcommand{\thefigure}{S\arabic{figure}}
\renewcommand{\thetable}{S\arabic{table}}

\captionsetup*{format=SIcaption}

\section{Biological relevance of transcription-induced active force in the model \label{sec:relevance}}

We assume that the active force is applied along the chromatin chain in order to model transcriptional elongation. 
Although modeling the complex process of transcription using mechanical force is  an over-simplification, we argue that it is reasonable. 
Single-molecule experiments have shown that RNA polymerase II (RNAPII) moves along DNA exerting large forces (up to 45 pN or more) during transcription \cite{Wang1998}. 
The helicase activity of RNAP, resulting in base pair opening, implies that the forces are exerted along the DNA bonds.  
Our simple active chromatin model provides insights into the dynamics of chromosomes in response to RNAP binding, which are difficult to obtain using experiments alone.  

The locus resolution, 1.2 kb $\sim$ 6 nucleosomes connected by linker DNAs, is smaller than the size of a typical gene, which is about 27 kb in human chromosomes \cite{Venter2001}.
At this resolution, the intergenic regions that correspond to inactive (B-type) loci are well separated from the gene-transcribing regions (A-type loci) in our model.
Each gene body, represented by multiple A-type loci, is presumed to contain a few actively transcribing RNAPII, as observed in the active burst phase of human genes \cite{Wan2021}. 

Is it reasonable to assume that the forces arising from the activity of RNAP in the gene-transcribing region propagate through other loci in the gene? 
We believe it is because the time scale for force propagation across the gene loci is expected to be less than the transcription time. 
Indeed, the time scale for force propagation in our monomer length scale is roughly 1 ns \emph{at most}, based on the Young's modulus and mass density of DNA, which is smaller than our simulation time unit ($\sim 70$ ns; see Methods in the main text). 
So the locally applied force by RNAP is felt outside the gene-transcribing regions (\emph{i.e.}, throughout a given A-type locus).

As described in the main text (see Fig.~\ref{fig:exp_model}D), the active force is exerted on each bond of the A-type loci in the form of a divergent or dipolar pair parallel to the bond vector, the magnitude of which is in accord with optical tweezer experiments \cite{Wang1998}.  
Such a force dipole should increase the distance between a bonded active locus pair. 
The increase in the bond distance is accompanied by an increase in the effective excluded volume of the A loci, potentially due to the local nucleosome motion enhanced by the tension forces propagated from actively transcribing RNAPs (see Fig.~\ref{fig:exp_model}D).
In the overdamped regime (as implemented in Eq.~\ref{eq:bd}), the propagation of $f_\text{RNAP} \approx 10$ pN would make the nucleosomes diffuse with the effective speed of $v=\frac{f_\text{RNAP}}{\zeta_\text{nuc}}\approx 5\,\text{nm}$ per simulation time step, where $\zeta_\text{nuc}$ is the friction coefficient for a nucleosome-sized particle in water. 
The locally activated motion of nucleosomes would increase the bond length between the monomers by inducing the volume exclusion.
The magnitude of the volume-excluding repulsion should be comparable to that of the tensional force on DNA exerted by RNAP. 
The typical estimate of $\approx$ 16 pN at $F=80$ used in our simulations is reasonable.

\section{Relevant time range of the experimental MSD data for comparison}

Nagashima \emph{et al.} tracked the motions of individual nucleosomes in live human RPE-1 cells and reported the mean square displacements (MSDs) of the nucleosomes with or without active transcription \cite{Nagashima2019}. 
They found that the MSD as a function of time reached a plateau within about 3 seconds, as shown in Fig.~\ref{fig:exp_model}A.
The saturation of the MSD curve is attributed to the constraint imposed on the nucleosome motion within the densely packed environment of chromatin.
In the MSD curve, we find that there are two different scaling regimes depending on the time scale.
For the control case (black color), the scaling exponent is about 0.5 for $t \le 0.5\,\text{s}$ whereas 0.23 for $t > 0.5\,\text{s}$.
The cases treated with the transcription inhibitors, $\alpha$-amanitin ($\alpha$-AM) and 5,6-dichloro-1-$\beta$-D-ribofuranosylbenzimidazole (DRB), show qualitatively the same scaling behavior.
Since the scaling behavior for $t \le 0.5\,\text{s}$ is more relevant to the polymer diffusion, we considered the data only for $t \le 0.5\,\text{s}$ to compare with our simulation results.

\section{Contributions of A- and B-type loci to dynamic properties}

\begin{figure}[h!]
\centering
\includegraphics[width = \textwidth]{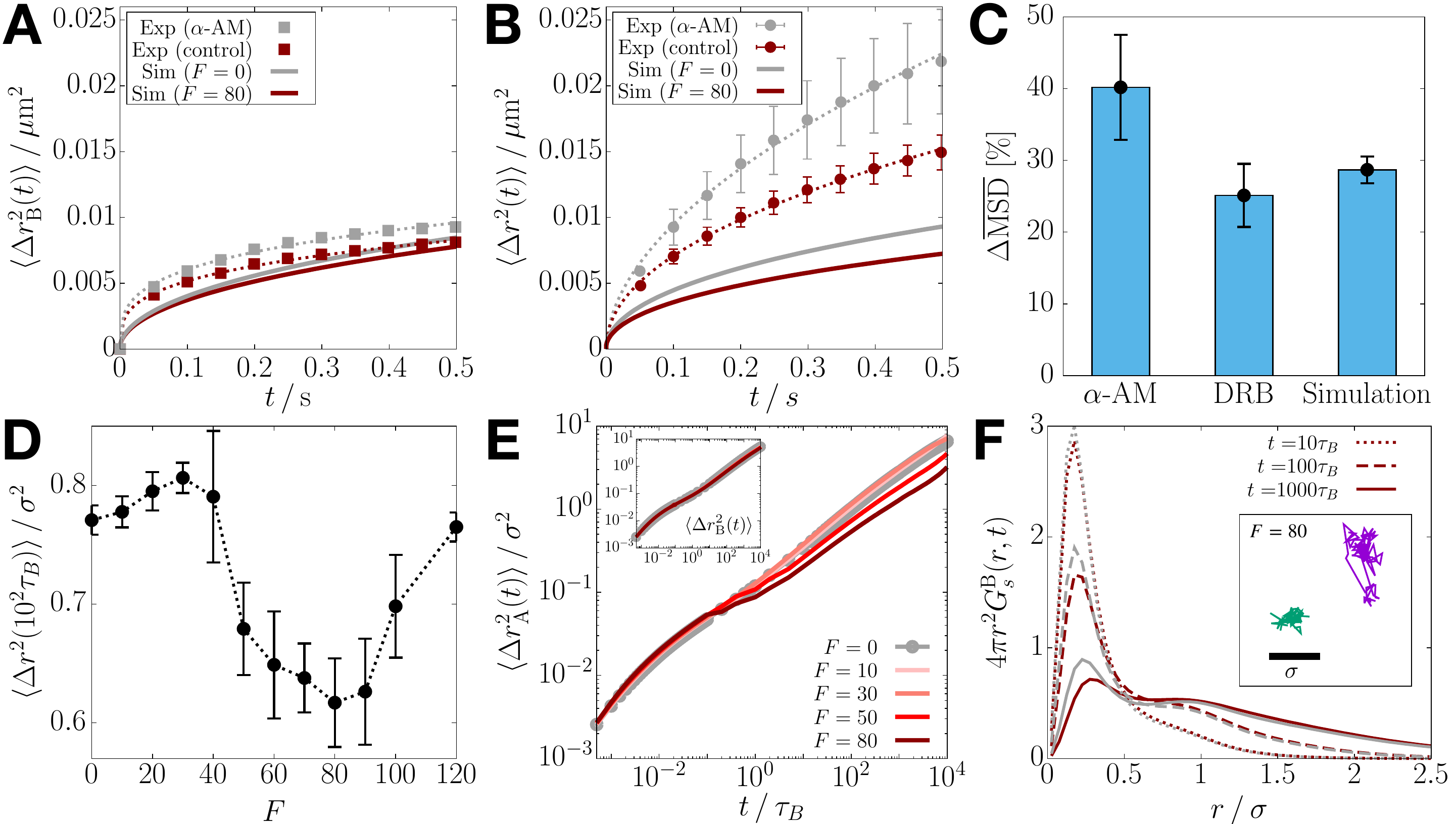}
\caption{Effect of active forces on the B-type loci dynamics is negligible. (A) Plot of $\langle \Delta r_\mathrm{B}^2(t) \rangle$ after treatment with $\alpha$-AM \cite{Nagashima2019} (squares), which we mimic by turning off the active force (solid lines). Dotted lines are the curves obtained by fitting the experimental data for the control (dark-red) and the inhibited (gray)
(B) Same as panel A, except for the plots correspond to overall MSD for all the loci. 
(C) Bar graphs comparing the increase in the overall MSD (as shown in panel B) relative to the control case when transcription is inhibited using $\alpha$-AM and DRB. Simulations mimic these cases using $F=0$ and $F=80$.
(D) Overall MSD at $t=100\tau_B$ as a function of $F$. The dotted lines are a guide to the eye. 
(E) Log-log plot of the MSDs for the A-type loci with different $F$. The inset shows the same plot for the B-type loci. 
(F) van Hove functions for the B loci with $F=0$ (gray) and $F=80$ (dark-gray) at $t = 10\tau_B$ (dotted line), $100\tau_B$ (dashed line), and $1000 \tau_B$ (solid line). The inset shows the 2-D projection of the trajectories of an active (green) and an inactive (purple) loci for $10^4\tau_B$ at $F=80$.
}
\label{fig:msd_supp}
\end{figure}

In the main text (Fig.~2B), we compared the difference between the passive and active cases of our simulations for a particular type of loci with the experimental results. 
Nagashima \emph{et al.} analyzed the MSDs based on the locations within the cell nucleus, specifying the data measured from either the interior or periphery of the nucleus \cite{Nagashima2019}.
We assumed that the experimental results for the interior (periphery) can be approximated as for euchromatin (heterochromatin).
For both A and B loci, the change in MSD upon treating the inhibitor is qualitatively comparable to the simulation results (Figs.~2A and \ref{fig:msd_supp}A). 
For a quantitative comparison, we calculated the relative increase in $\langle \Delta r_\mathrm{A}^2(t) \rangle$ and $\langle \Delta r_\mathrm{B}^2(t) \rangle$ upon turning off the activity, using 
\begin{equation}
\Delta\overline{\mathrm{MSD}}_\text{exp}^{\mu} = \frac{1}{t_\text{diff}}\int_0^{t_\text{diff}} dt \frac{\langle \Delta r_\mu^2(t)\rangle_\text{inhibited} - \langle \Delta r_\mu^2(t)\rangle_\text{control}}{\langle\Delta r_\mu^2(t)\rangle_\text{control}}~,
\label{eq:delta_msd_exp}
\end{equation}
for the experiment and 
\begin{equation}
\Delta\overline{\mathrm{MSD}}_\text{sim}^\mu = \frac{1}{t_\text{diff}}\int_0^{t_\text{diff}} dt \frac{\langle \Delta r_\mu^2(t)\rangle_{F=0} - \langle \Delta r_\mu^2(t)\rangle_{F=80}}{\langle\Delta r_\mu^2(t)\rangle_{F=80}}~,
\label{eq:delta_msd_sim}
\end{equation}
for the simulation, where $t_\text{diff} = 0.5\,\text{s}$.

\begin{figure}[h!]
\centering
\includegraphics[width = \textwidth]{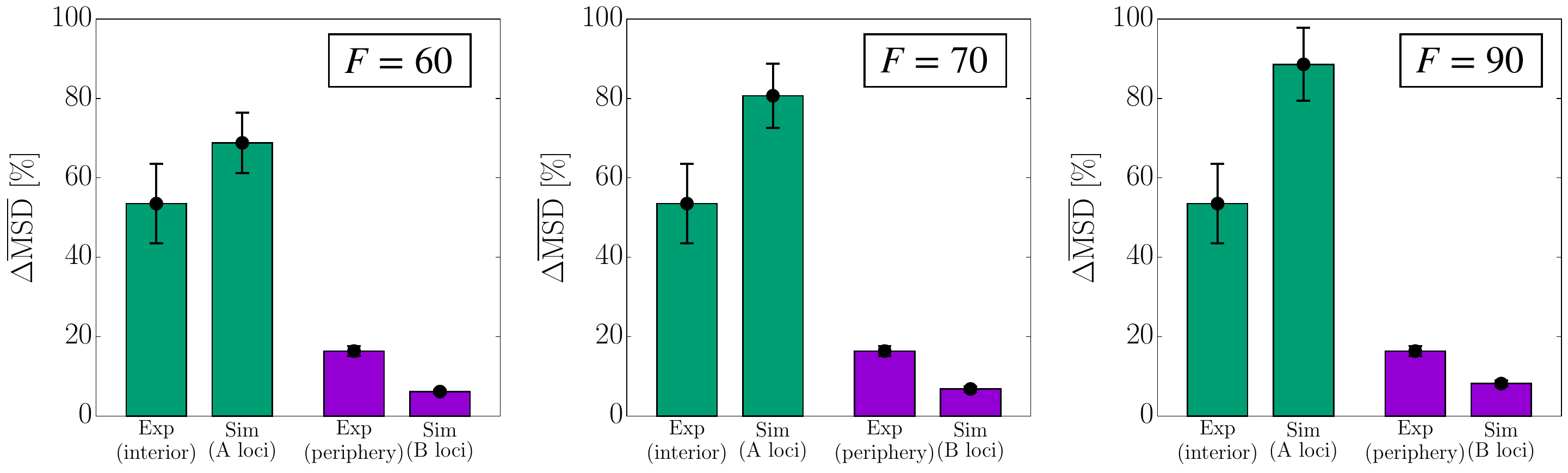}
\caption{Simulations in the intermediate range of $F$, $60 \le F \le 90$, give  results that are in a good agreement with experiments. Bar graphs comparing $\Delta \overline{\text{MSD}}$ for the A and B loci separately [Eqs.~(\ref{eq:delta_msd_exp})-(\ref{eq:delta_msd_sim})] between the experiment and simulation results with $F=60$ (left), 70 (center), and 90 (right). The results for $F=80$ are given in the main text (Fig.~2B). Taken together, these results show that over the range, $60 \le F \le 90$, the agreement with experiments is reasonable, which further reinforces the robustness of the proposed mobility inhibition mechanism.
}
\label{fig:msd_cmp_F60-90}
\end{figure}

We found that the comparison between $\Delta\overline{\mathrm{MSD}}_\text{exp}$ and $\Delta\overline{\mathrm{MSD}}_\text{sim}$ is more quantitative for the B loci than the A loci (Fig.~2B, main text).
The quantitative difference could arise because every interior measurement does not necessarily correspond to the euchromatin whereas the periphery measurements are mostly for the heterochromatin. 
It is notable that the comparison between simulations and experiments is qualitatively similar for $F$ in the range, $60 \le F \le 90$ (Fig.~\ref{fig:msd_cmp_F60-90}).
If we compare the overall MSD for all the loci, $\langle\Delta r^2(t)\rangle = [N_\text{A} \langle\Delta r_\text{A}^2(t)\rangle + N_\text{B} \langle\Delta r_\text{A}^2(t)\rangle]/N$ with the experimental results (Fig.~\ref{fig:msd_supp}B), we find that $\Delta\overline{\mathrm{MSD}}_\text{sim}$ quantitatively agrees with $\Delta\overline{\mathrm{MSD}}_\text{exp}$ (Fig.~\ref{fig:msd_supp}C).
Note that the overall MSD, $\langle\Delta r^2(t)\rangle$, at a given lag time (\emph{i.e.}, $t=100\tau_B \approx 0.07\,\text{s}$) reflects the non-monotonic change of $\langle\Delta r_\text{A}^2(t)\rangle$ upon increasing $F$ from 0 to 120 (Fig.~\ref{fig:msd_supp}D).

\begin{figure}[h!]
\centering
\includegraphics[width = \textwidth]{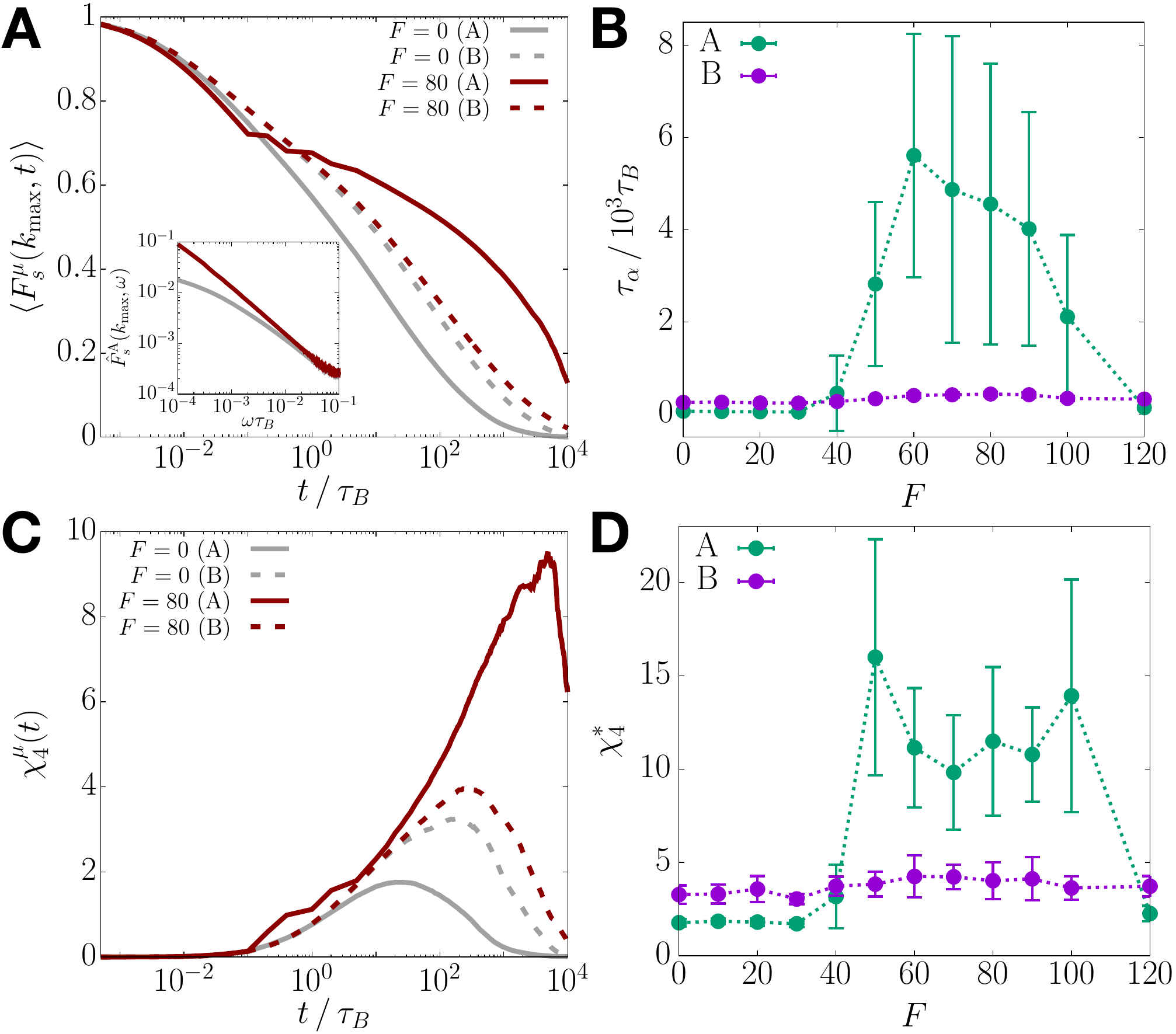}
\caption{Slow dynamics and heterogeneous mobilities are associated with the euchromatin loci. 
(A) Comparison between $\langle F_s^\mathrm{A} (k_\mathrm{max},t) \rangle$ and $\langle F_s^\mathrm{B} (k_\mathrm{max},t) \rangle$ (Eq.~\ref{eq:fskt_AB}), shown by solid and dashed lines, respectively, for $F=0$ and $F=80$. The inset shows the log-scale plot of $\hat{F}_s^\mathrm{A} (k_\mathrm{max},\omega)$, the frequency-domain Fourier transform of $\langle F_s^\mathrm{A} (k_\mathrm{max},t) \rangle$.
(B) The relaxation time of A and B loci for different $F$ values. 
(C) Comparison between $\chi_4^\mathrm{A}(t)$ and $\chi_4^\mathrm{B}(t)$ (Eq.~\ref{eq:chi4_AB}), shown by solid and dashed lines, respectively, for $F=0$ and $F=80$. 
(D) The maximum value in $\chi_4^\mathrm{A}(t)$ and $\chi_4^\mathrm{B}(t)$ for different $F$. The dotted lines in panels B and D are a guide to the eye. 
}
\label{fig:fskt_AB}
\end{figure}

We also decomposed $\langle F_s (k_\mathrm{max},t) \rangle$ and $\chi_4$, shown in Fig.~3 of the main text, into contributions from the A and B loci separately. 
For the $\mu$-type loci, we define 
\begin{equation}
F_s^\mu (|\mathbf{k}|,t) = \frac{1}{N_\mu} \sum_{i=1}^N \delta_{\nu(i)\mu}e^{i \mathbf{k}\cdot[\rp_j(t) - \rp_j(0)]}~,
\label{eq:fskt_AB}
\end{equation}
and
\begin{equation}
\chi_4^\mu(t) = N_\mu\left[\left<F_s^\mu (k_\text{max},t)^2 \right> - \left< F_s^\mu (k_\text{max},t) \right> ^2 \right]~,
\label{eq:chi4_AB}
\end{equation}
where $k_\text{max} = 2\pi/\sigma$.
Figure \ref{fig:fskt_AB}A compares $\langle F_s^\mathrm{A} (k_\mathrm{max},t) \rangle$ and $\langle F_s^\mathrm{B} (k_\mathrm{max},t) \rangle$. 
For $F=0$,  $\langle F_s^\mathrm{A} (k_\mathrm{max},t) \rangle$ decays faster than $\langle F_s^\mathrm{B} (k_\mathrm{max},t) \rangle$, whereas $\langle F_s^\mathrm{A} (k_\mathrm{max},t) \rangle$ decays slower at $F=80$. 
Remarkably, $\langle F_s^\mathrm{A} (k_\mathrm{max},t) \rangle$ at $F=80$ exhibits a shoulder prior to the structural relaxation. 
This shoulder, starting near $\langle F_s^\mathrm{A} (k_\mathrm{max},t) \rangle =0.7$ and $t<\tau_B$, arises due to the enhanced caging by neighboring particles around the probe, which results from the collective solid-like ordering. The Fourier transform of the scattering function (Fig.~\ref{fig:fskt_AB}A, inset) illustrates the increase in the  slowly-decaying low-frequency modes, which correspond to the collective motions of the ordered particles, whereas the fast-decaying high-frequency modes are rather random motions.
The similar trend is also reflected in $\chi_4^\mathrm{A}(t)$ and $\chi_4^\mathrm{B}(t)$, as shown in Fig.~\ref{fig:fskt_AB}C. 
At $F=0$, the peak of $\chi_4^\mathrm{A}(t)$ appears earlier than $\chi_4^\mathrm{B}(t)$, albeit with a smaller amplitude. 
In contrast, at $F=80$, $\chi_4^\mathrm{A}(t)$ exhibits a larger peak which appears at a later time than $\chi_4^\mathrm{B}(t)$, suggestive of large-scale coherent motions. 
In Fig.~\ref{fig:fskt_AB}B, the plots of $\tau_\alpha^\mathrm{A}$ and $\tau_\alpha^\mathrm{B}$, the relaxation times determined from $\langle F_s^\mathrm{A} (k_\mathrm{max},t) \rangle$ and $\langle F_s^\mathrm{B} (k_\mathrm{max},t) \rangle$, respectively, indicate that the retardation of the active polymer chain dynamics compared to the passive case is predominantly driven by the changes in the dynamics of the A loci. 
In accord with this observation, the increase in dynamic heterogeneity at $F > 0$ is also originated from the A loci rather than the B loci (Fig.~\ref{fig:fskt_AB}D). 

\subsection*{Alignment of the sampled configurations}

When analyzing the simulation data to compute the dynamic properties, we align the whole polymer with respect to the initial configuration to remove the global translational and rotational drifts. 
Specifically, at each time frame in a given trajectory, we shift the polymer configuration to minimize the root mean square displacement from the initial configuration. 
As a result, $\langle{\Delta r^2_\mu}(t)\rangle$ measures the internal diffusion of individual loci in the reference frame of the center of mass of the system.

\section{Bending angle distribution}

The extensile active force applied to each A-A bond could induce a net perpendicular force to the chain contour for the A loci, as implied in Fig.~\ref{fig:exp_model}E in the main text.  
The net force on the $i^\text{th}$ locus, given in Eq.~\ref{eq:f_a} if the $(i-1)^\text{th}$ and $(i+1)^\text{th}$ are also A-type, becomes $\mathbf{f}_i(t)  = - f_0 \bm{\kappa}_i(t)$ where 
$\bm{\kappa}_i(t) = \mathbf{\hat{b}}_{i}(t) - \mathbf{\hat{b}}_{i-1}(t)$ is the curvature vector at the $i^\text{th}$ point of the discrete chain connected by the normal bond vectors.  
Hence, the applied force is should decrease the local bending angles for the A-type loci.
To probe the effect of the force on the curvature, we calculated the distribution of the angle formed between two consecutive A-A bonds, defined as,
\begin{equation}
P(\theta_\text{A}) = \frac{\sum_{i=2}^{N-1} \delta_{\nu(i-1)\text{A}}\delta_{\nu(i)\text{A}}\delta_{\nu(i+1)\text{A}}\langle\delta(\cos^{-1}(\mathbf{\hat{b}}_i \cdot\mathbf{\hat{b}}_{i+1}) -  180^\circ + \theta_\text{A})\rangle}{\sin\theta_\text{A} \sum_{i=2}^{N-1} \delta_{\nu(i-1)\text{A}}\delta_{\nu(i)\text{A}}\delta_{\nu(i+1)\text{A}}}~.
\label{eq:bend_angle}
\end{equation}
Upon increasing the activity, the distribution becomes broader as the peak at $\theta_\text{A} = 71^\circ$ shifts to the left (Fig.~\ref{fig:bend_angle}, top), indicating that the bending angle tends to decrease due to the active force. 
Remarkably, $P(\theta_\text{A})$ at $F=80$ shows more pronounced peaks at $\theta_\text{A} = 60^\circ$, $120^\circ$, and $180^\circ$, along with another one at $\theta_\text{A} =90^\circ$, which reflect the structural transition involving the transient FCC-like order.  
The  less pronounced peak near $\theta_\text{A} = 147^\circ$ is due to the hexagonal closed packing (HCP) geometry that arises at the surface of the polymer globule \cite{Ni2013}.
In contrast, the bending angle distribution for the B loci, $P(\theta_\text{B})$, similarly defined as in Eq.~\ref{eq:bend_angle}, does not change upon increasing $F$ (Fig.~\ref{fig:bend_angle}, bottom).

\section{Correlation time of bond-orientational order parameter\label{sec:transient}}

We calculated the temporal correlation in $q_{12}$ for a given locus type by using the time correlation function, 
\begin{equation}
C_{q_{12}}^\mu (t) =  \frac{\langle \delta q_{12}(t)\delta q_{12}(0)\rangle_\mu}{\langle (\delta q_{12})^2 \rangle_\mu} = \frac{1}{\langle (\delta q_{12})^2 \rangle_\mu } \left[\frac{1}{N_\mu}\sum_{i=1}^N \delta_{\nu(i)\mu}\langle q_{12}(i;t) q_{12}(i;0) \rangle - \langle q_{12}\rangle_\mu^2 \right]~.
\label{eq:q_corr}
\end{equation}
In Fig.~\ref{fig:transient}B, $C_{q_{12}}^\text{A} (t)$ and $C_{q_{12}}^\text{B} (t)$ are shown for five independent simulation trajectories at $F=80$. 
Due to activity-induced ordering in the A loci, $C_{q_{12}}^\text{A} (t)$ decays over longer time scale than $C_{q_{12}}^\text{B} (t)$.
Nevertheless, a majority of the order decorrelates in less than $0.2\tau_B$ (Fig.~\ref{fig:transient}B, inset).
The maximum time over which $C_{q_{12}}^\text{A} (t)$ decays is $1,000\tau_B\approx 0.7\,s$.
The transient nature of the order is also visually discernible from a trajectory at $F=80$, if we mark the loci whose $q_{12}$ is larger than a threshold at a given time. 
The threshold is chosen to be $q_{12} = 0.45$, which is approximately the midpoint between the liquid ($q_{12}\approx0.3$) and the FCC solid ($q_{12}\approx0.6$).
The simulation snapshot with this labeling illustrates that only subpopulations of the A loci exhibit the order (Fig.~\ref{fig:transient}C, left). 
This ordering fluctuates from trajectory to trajectory. 
In contrast, the quenched active polymer shows a larger cluster of FCC-like ordering that stays indefinitely (Fig.~\ref{fig:transient}C, right).
The movies for the trajectories including the snapshots shown in Fig.~\ref{fig:transient}C are given in Movies S1 and S2.

\section{FCC-like ordering does not alter the chromosome conformation\label{sec:6}}

Does the FCC-like order induced by the active force affect the structure of the folded chromosome? To answer this question, we computed $P(s)$, the average contact probability at a given genomic pair distance, $s$, using
\begin{equation}
P(s) =  \frac{1}{N - s} \sum_{i=1}^{N-1} \sum_{j=i+1}^{N} \delta_{|i-j|,s}C_{i,j} ~,
\label{eq:pofs}
\end{equation}
where $C_{i,j}$ is the contact matrix from either the Hi-C experiment or the simulation. 
In our previous study \cite{Shi2018a}, we demonstrated that the CCM quantitatively reproduces the $s$-dependence of the pair contact probability, especially the power-law scaling exponent in $P(s)$. 
In Fig.~\ref{fig:pofs}A, $P(s)$ for the CCM with $F=80$ is nearly identical to that for $F=0$, showing  reasonable agreement with the Hi-C result by capturing the two distinct scaling exponents in the power-law decay ($\sim s^{-0.8}$ for $s \lesssim 0.5$ Mbps and $\sim s^{-1.5}$ for $s > 0.5$ Mbps).
We also calculated the average contact probability for a given pair type, using
\begin{equation}
P_{\mu\gamma}(s) =  \frac{\sum_{i=1}^{N-1} \sum_{j=i+1}^{N} \delta_{\nu(i)\mu}\delta_{\nu(j)\gamma}\delta_{|i-j|,s}C_{i,j}}{\sum_{i=1}^{N-1} \sum_{j=i+1}^{N} \delta_{\nu(i)\mu}\delta_{\nu(j)\gamma}\delta_{|i-j|,s}} ~,
\label{eq:pofs_type}
\end{equation}
where $\mu$ and $\gamma$ is either A or B. As shown in Fig.~\ref{fig:pofs}B, there is no significant change in the $s$-dependence of $P_{\mu\gamma}(s)$ upon increasing $F$ from 0 to 80.

For a further comparison between the 3D structures, we plotted the mean distance matrices, $\bar{r}_{i,j}$, at $F=0$ and $F=80$ in Fig.~\ref{fig:pofs}C.
Here, the bar in $\bar{r}_{i,j}$ denotes the average over a single trajectory, and thus we compare the averaged 3D structures that are propagated from the same initial configuration but with different $F$. 
The visual inspection suggests that there is no significant difference between the distance matrices. 
The evaluation of the relative mean absolute error (RMAE), defined as 
\begin{equation}
\mathrm{RMAE} =  \frac{2}{N (N+1)}\sum_{i=1}^N \sum_{j=i}^{N} \frac{\bar{r}_{i,j} (F=0) -  \bar{r}_{i,j} (F=80)}{\bar{r}_{i,j} (F=0)}~,
\label{eq:error}
\end{equation}
yields $\sim10$\%, which also indicates quantitative similarity.
In Fig.~\ref{fig:pofs}D, the calculated and Hi-C contact maps are compared.  The simulated contact maps, both at $F=0$ and $F=80$, faithfully capture the large-scale patterns in the Hi-C contact map (Fig.~\ref{fig:pofs}D), although there are quantitative deviations in the A/B segregation on small length scale (Fig.~\ref{fig:pofs}E).
Such deviations could arise mainly because of the minimal features of our polymer model as well as the finite-size effect (see the caption of Fig.~\ref{fig:pofs}E for more detail). 
Nevertheless, the predicted contact map at $F=80$ is essentially the same as the one at $F=0$.
Therefore, the activity-induced ordering does not alter the overall chromosome organization significantly.

\section{Dynamic properties of a chromosome 10 segment\label{sec:7}}
We also simulated a 4.8-Mbp segment of chromosome 10 (Chr10), which has more A-type loci than the 4.8-Mbp segment of Chr5 (Fig.~\ref{fig:chr10}A); $N_\text{A}/N = 0.58$ for Chr10 whereas 0.25 for Chr5.  
We calculated the dynamical properties from the simulations with active force to compare with the findings for Chr5 presented in the main text.
In Fig. \ref{fig:chr10}B, the MSDs for different $F$ values are shown in log scale. 
The dynamics becomes slower upon increasing $F$ and the mobility suppression is maximized at $F=80$, which follows the same trend found for Chr5 (Fig.~\ref{fig:chr10}C).
Notably, the extent of the suppression is larger while the scaling exponent of the MSD for $F=80$ is smaller than for Chr5 ($\alpha = 0.40$ versus 0.43).
Similarly, the plots of $\langle F_s(k_\text{max},t) \rangle$ and $\chi_4(t)$ show the trends that are consistent with those for Chr5 (Figs.~\ref{fig:chr10}D and \ref{fig:chr10}E); the relaxation is slower and the dynamic heterogeneity is larger than in Chr5. The change due to active force in Chr10 is larger than Chr5, which is a consequence of the higher fraction of A-type loci in Chr10 compared to Chr5.
The radial distribution function for A-A pairs, $g_\text{AA}(r)$, and the distribution of the BOO parameter for the A loci, $P_\text{A}(q_{12})$, show the same trends as those for Chr5 shown in Figs.~\ref{fig:struc_main}A and \ref{fig:struc_main}B of the main text. 
Overall, the qualitative results are similar between Chr5 and Chr10, which shows that ordering of the active loci for a range of $F$ is a robust finding.

\section{Random copolymer with $F\neq0$ does not exhibit fluid-to-FCC transition\label{sec:8}}

We also investigated the effect of $F$ on a copolymer chain with random sequence by shuffling the loci in Chr5 without changing the number of A or B type loci. 
We performed Brownian dynamics simulations with the active force applied in the same manner as in the wild type (WT) Chr5. 
This randomized chain does not undergo microphase separation like the WT sequence, whose A/B sequence reflects the genetic activity of an actual chromosome (Chr5 and Chr10). 
As the favorable interactions between the monomers of the same type is less likely in a   random sequence, the dynamics of the randomized chain is moderately faster than the WT, as shown in Fig.~\ref{fig:rand_seq}A.
Nonetheless, the dynamics is glass-like (Figs.~\ref{fig:rand_seq}B-\ref{fig:rand_seq}C), as shown previously \cite{Shi2018a}.
Upon increase of $F$, as shown in Fig.~\ref{fig:rand_seq_main}A, the MSD for the random sequence decreases by less than 2\% only if $F$ exceeds 80. 
The decomposition of the MSD into A and B contributions shows that the slightly reduced dynamics is driven by the B-type monomers (Fig.~\ref{fig:rand_seq_main}A). 
In contrast, the movement of the A-type is enhanced due to the active force. 
The distributions of the BOO parameters, $P_\text{A}(q_{12})$ and $P_\text{B}(q_{12})$, confirm that the reduced dynamics for $F > 80$ is not accompanied by the FCC-lattice formation, as found in the WT sequence (Figs.~\ref{fig:rand_seq_main}B and \ref{fig:rand_seq}D). 

\section{Effect of removal of loop anchors on the dynamics\label{sec:9}}

We tested if the CTCF-mediated loops contribute to the change in the active loci dynamics observed when the extent of activity is varied. 
We ran Brownian dynamics simulations of the single CCM chain (Chr5) with all the loops removed (\emph{i.e.}, there is no harmonic potential applied between the locus pairs identified as the CTCF loop anchors). 
It should be noted that CTCF loop loss eliminates TAD-like structures but enhances compartment formation (microphase separation between A and B type loci). Other simulation details are the same as described above. 

For the CCM without the loops, the overall MSD also decreases to almost the same extent as with the loops, upon increasing $F$ from 0 to 80 (Fig.~\ref{fig:loop}A). 
There is a modest increase in the MSD at $F=0$ compared to the case with the loops, which is consistent with the experimental observation for the dynamic change upon the CTCF-cohesin loop loss \cite{Nozaki2017}.
The MSD for the A-type loci decreases modestly, with the scaling exponent decreasing from  from 0.38 to 0.36 (Fig.~\ref{fig:loop}B).
The change in the MSD for the B-type loci upon removing the loops is relatively minor (Fig.~\ref{fig:loop}C).
We also found that the change in $\langle F_s^\mu (k_\mathrm{max},t) \rangle$ and $\chi_4^\mu(t)$ due to the loop disruption is not significant  either at $F=0$ or $F=80$ (Figs.~\ref{fig:loop}D-\ref{fig:loop}E).
Interestingly, although the loop removal leads to more suppressed A-loci dynamics at $F=80$, the distribution of $q_{12}$ for the A-type loci is nearly identical to that with the loops (Fig.~\ref{fig:loop}F). 
The conservation of $P_\text{A}(q_{12})$  suggests that the transition between disordered and ordered states occurs more frequently, which makes long-time scale motions slower, while the coexistence point remains the same at a given level activity. 
Therefore, the contribution of the loop constraints to the A-loci dynamics is not so pronounced as that of the morphological transition induced by the active force.

\section{Effects of variations in non-bonding interactions\label{sec:eps_nb}}

As specified in a previous section, we applied the condition of $\epsilon_\text{AA} = \epsilon_\text{BB}$ to the non-bonding interaction parameters for the CCM simulations. 
This simple assumption was sufficient to demonstrate the CCM's capability of capturing experimental results faithfully \cite{Shi2018a}. 
However, this assumption may not always hold because heterochromatin (B-type) is considered to have more compact nature than euchromatin (A-type). 
To test if the differentiated A-A and B-B interactions affect our results, we performed additional simulations for the Chr5 segment with $\epsilon_\text{AA} < \epsilon_\text{BB}$, which represents a larger extent of compaction for the B-type loci than the A-type. 
Specifically, we set $\epsilon_\text{AA} = 2.4k_BT$ and $\epsilon_\text{BB} = 2.6k_BT$, whereas we used  
the same A-B interaction parameter, $\epsilon_\text{AB} = \frac{9}{11}\epsilon_\text{AA} = 1.96 k_B T$. 
These interaction parameters still ensure  microphase separation between A and B loci, as determined by  the Flory-Huggins theory \cite{Huggins1941,Flory1941}.
Other simulation details are the same as described above. 

In Fig.~\ref{fig:eps_diff}A, the MSD for the A-type loci decreases with $F=80$ and still captures the experimental results qualitatively (\emph{cf.} Fig.~2A, main text). In Fig.~\ref{fig:eps_diff}B, the MSD for the A-type at $t= 10^3 \tau_B \approx 0.7s$ shows the dynamic re-entrance with respect to $F$ whereas the change in the MSD for the B-type is relatively insignificant, which is similar to the result  in Fig.~\ref{fig:msd}C. 
In addition, the structural quantities, $g_\text{AA}(r)$ (Fig.~\ref{fig:eps_diff}C) and $P_\text{A}(q_{12})$ (Fig.~\ref{fig:eps_diff}D), do not change from the original results shown in Figs.~\ref{fig:struc_main}A and \ref{fig:struc_main}B. 
The FCC-like arrangement of the active loci is visually noticeable from the simulation snapshot in Fig.~\ref{fig:eps_diff}C. 
Therefore, the $F$-induced dynamical slowdown and transient ordering of the A loci are not affected by increasing $\epsilon_\text{BB}$ relative to $\epsilon_\text{AA}$. 
We surmise that the most important conclusion that the locus dynamics is suppressed during transcription, which is in accord with experiments, is robust as shown in Fig.~\ref{fig:eps_diff}.

\section{Comparisons with a previous study \cite{Saintillan2018}\label{sec:compare}}
Saintillan, Shelley, and Zidovska (SSZ)  studied the effect of extensile active force on the chromosome dynamics using a polymer model \cite{Saintillan2018}. 
Although both the present work and the SSZ study show that active forces induce order in chromatin, there are a few differences as well. 
(1) In the SSZ model the bond length is fixed.  
Here, it is the activity-induced (in a narrow range of $F$) increase in the bond length that produces transient solid-like ordering by making the bond length commensurate with the non-bonding attractive interaction (see  the main text).
In the SSZ study, nematic order, which emerges in the presence of activity,  
increases monotonically as $F$ increases, whereas in our model the emergence of solid-like order is non-monotonic. 
(2) The SSZ simulations were performed by confining the polymer chain, which likely produces alignment.  
Our simulations do not confine the single chain, and hence the mechanism of FCC-like ordering is  different from the formation of  nematic phase. 
(3) Activity in the SSZ arises due to fluid flow that is coupled to motion of the beads on the chain, which is an insightful model for probing the effect of active forces on a polymer. In the present work, we envision that RNAPII exerts forces during transcription.
(4) In the unconfined case, the SSZ model predicts coil-stretch transition in the absence of activity (see Fig. S1 in SSZ).  In our model, at low $F$ the dynamics exhibits glass-like behavior with no ordering. Despite these differences the formation of ordered structures promoted by active forces is a common theme in both the studies.

\section{Effect of gene co-activation on the dynamics \label{sec:g1_l2}}

A recent study \cite{Bohrer2023} showed that there is a significant correlation between spatial proximity of genes and their co-expression.  
Our study provides some insight into the underlying mechanism for the spatial coupling of co-activated genes since active forces are applied to all the active genes in the simulated Chr5 region. 
It is worth examining  the dynamical consequences when only one gene is active and others are turned off.  To address this issue, we repeated the simulations by applying the active forces only to the loci corresponding to a particular single gene. 
We considered the \emph{TCOF1} gene, which is located at 149,737,202--149,779,871 bp on Chr5 (see Fig.~\ref{fig:g1_l2}A).
The \emph{TCOF1} gene is about 43 kb long and corresponds to 37 loci ($3223 \le i \le 3259$) in the CCM polymer chain ($N=4000$). Because this gene is not insulated by the CTCF loops, it is potentially less affected by the loops. 
Application of the active force on a single gene results in  enhanced mobility of the gene loci (Fig.~\ref{fig:g1_l2}B), whereas the overall MSD for all the A loci does not show a noticeable change (Fig.~\ref{fig:g1_l2}B, inset). 
We  repeated the simulations by exerting the active forces only to the first two consecutive A-A bonds in the gene, \emph{i.e.}, three A-type loci ($i=3223,3224,3225$).
The active forces on the three loci give some changes in the MSD of those loci, but the variations are so large that the overall change upon increasing the activity is not significant (Figs.~\ref{fig:g1_l2}C and \ref{fig:g1_l2}D). 
Fig.~\ref{fig:g1_l2}D shows that the MSD for the gene loci increases monotonically as $F$ is increased when only one gene is transcribed. 
Therefore, it is reasonable to suggest that activity-induced mobility suppression requires co-activation of genes in the given chromosome region.   

\clearpage

\begin{figure}[h!]
\centering
\includegraphics[width = 5.6 in]{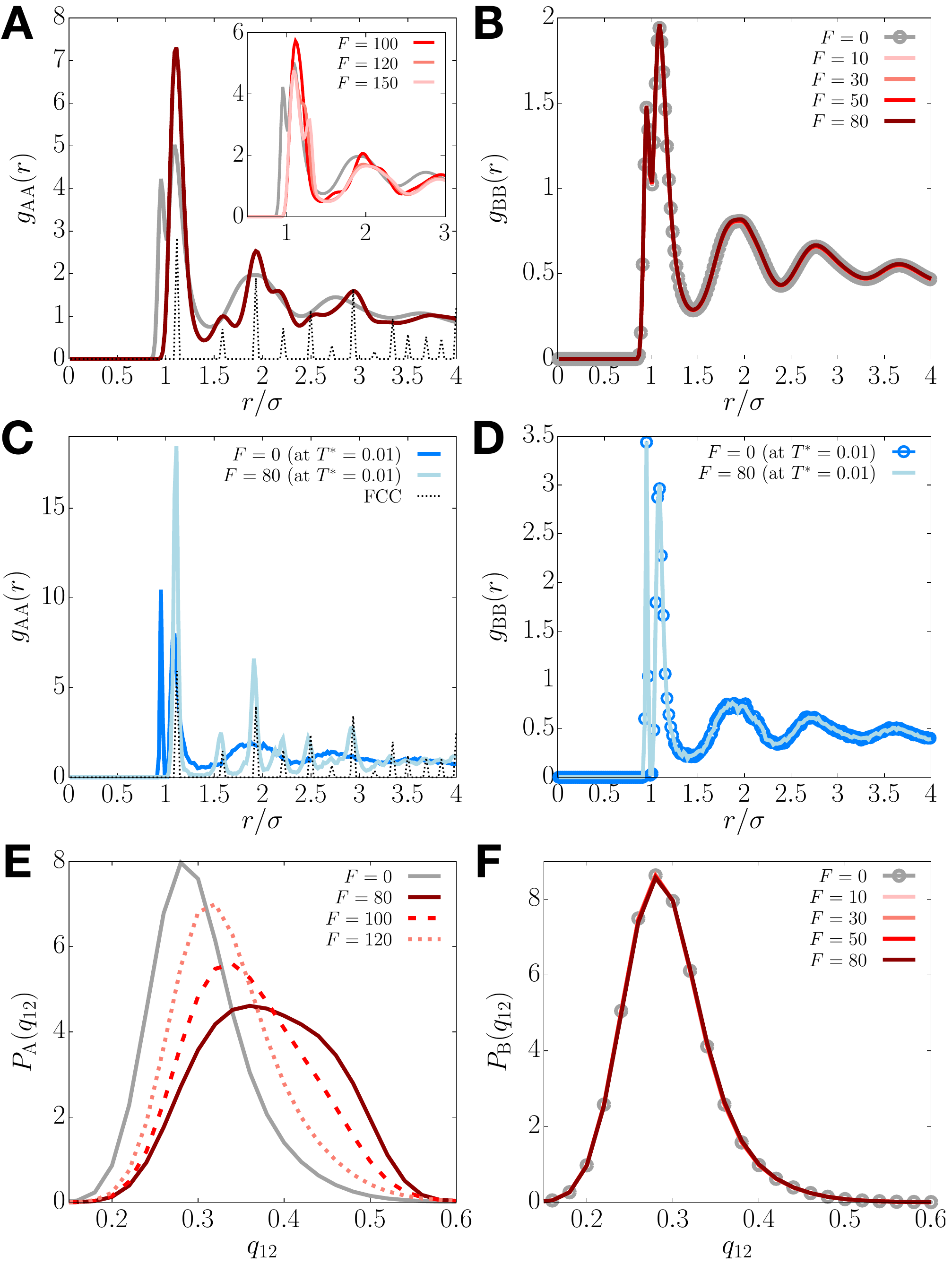}
\caption{$F$-dependent structural properties for A and B loci. 
(A) Radial distribution function for A-A loci pairs, $g_\text{AA}(r)$, at $F=0$ (gray), $F=80$ (maroon), and $F > 80$ (inset). $g(r)$ for a FCC crystal is shown with the dotted line for comparison (scaled arbitrarily). The inset shows that the peaks associated with the FCC phase disappear when $F$ exceeds 100. The peak at $r \lesssim \sigma$ for $F=0$ corresponds to the bonded pairs ($i^\text{th}$ and $(i+1)^\text{th}$ loci), whose location shifts to the right as $F$ increases.
(B) Plots of $g_\text{BB}(r)$ for different $F$. $g_\text{BB}(r)$ is independent of $F$ and shows the behavior that is reminiscent of a dense fluid at all $F$ values, including $F=0$.
(C) Inherent structure $g_\text{AA}(r)$ for $F=0$ (blue) and $F=80$ (light blue).
(D) Same as panel C except the results for the B-type loci. 
(E) Distributions of the BOO parameter for the A-type loci [Eqs.~(\ref{eq:bo_1})-(\ref{eq:q_dist})] for $F\ge80$ and $F=0$. 
(F) Same as panel E, except the results are for the B-type loci. Like $g_\text{BB}(r)$, $P_\text{B}(q_{12})$ is independent of $F$.
}
\label{fig:gofr_BOO_supp}
\end{figure}

\begin{figure}[h!]
\centering
\includegraphics[width = 5.6in]{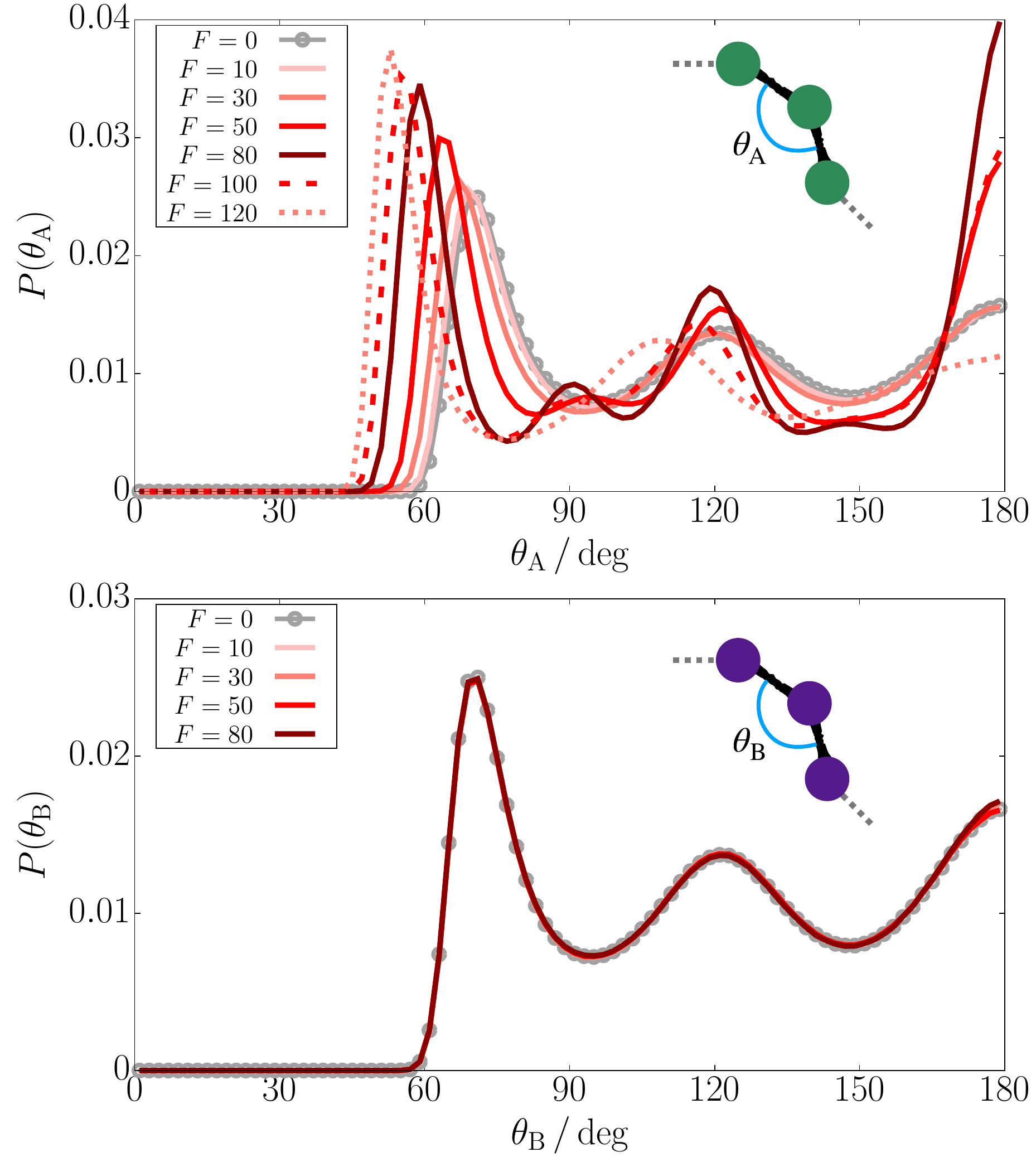}
\caption{Bending angle distributions for three consecutive A (top) and B (bottom) loci (Eq.~\ref{eq:bend_angle}) at different $F$. 
}
\label{fig:bend_angle}
\end{figure}

\begin{figure}[h!]
\centering
\includegraphics[width = 5.6in]{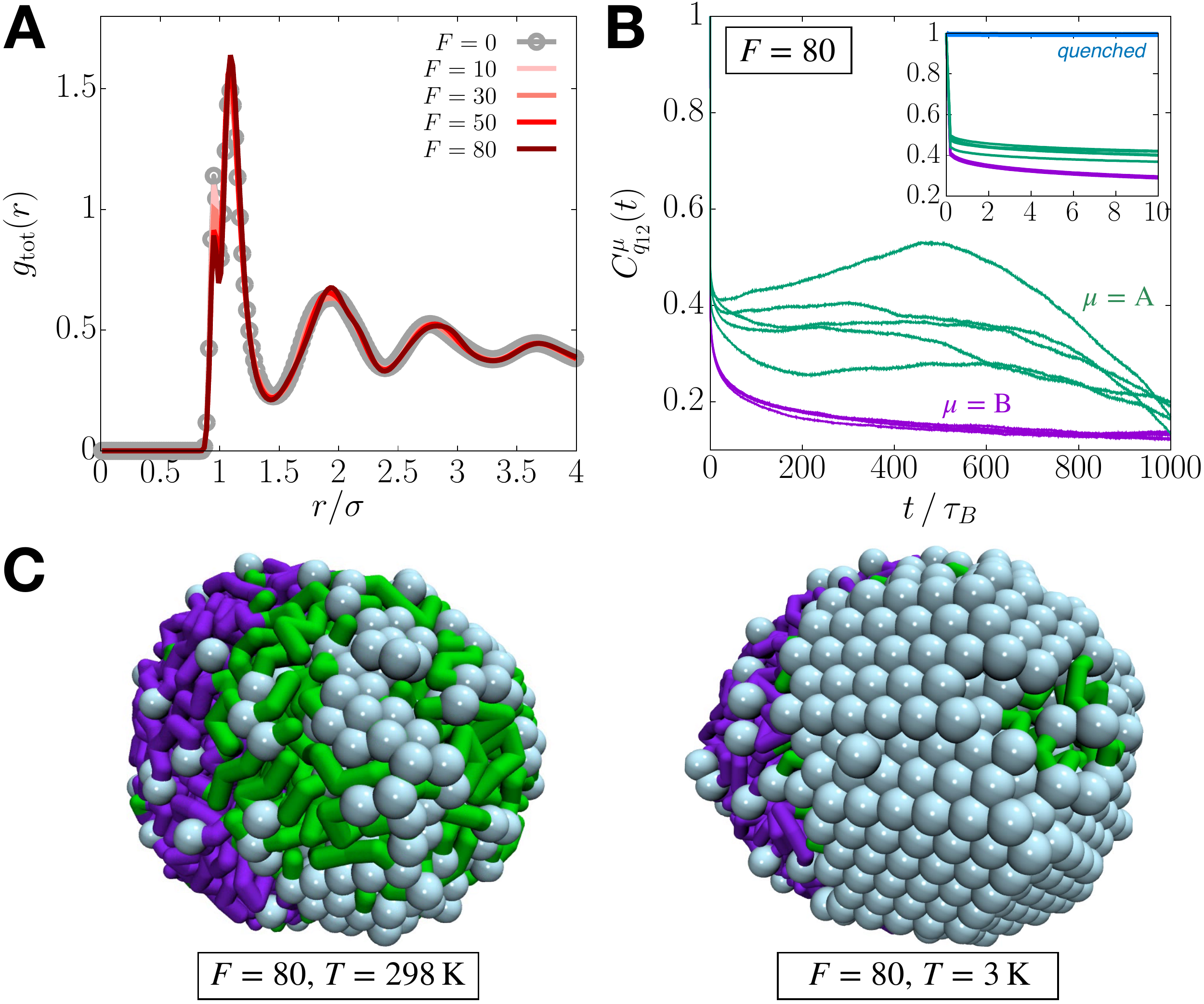}
\caption{Activity-induced order in the A loci exists only transiently.
(A) Radial distribution function for all the locus pairs at different $F$.
(B) Plots of the time correlation functions, $C_{q_{12}}^\text{A} (t)$ and $C_{q_{12}}^\text{B} (t)$ (Eq.~\ref{eq:q_corr}), for five independent trajectories at $F=80$.  The inset shows the short-time changes in the correlation functions, where $C_{q_{12}}^\text{A} (t)$ for the quenched polymer is shown in blue. 
(C) Simulation snapshots for the active copolymer with $F=80$ at room temperature (left) and the one quenched to low temperature (right). Green and purple colors represent the A and B loci, respectively. The light blue spheres indicate the loci with $q_{12}(i) > 0.45$. The movies for these simulation trajectories are given in Movies S1 (left) and S2 (right).
}
\label{fig:transient}
\end{figure}

\begin{figure}[h!]
\centering
\includegraphics[width = \textwidth]{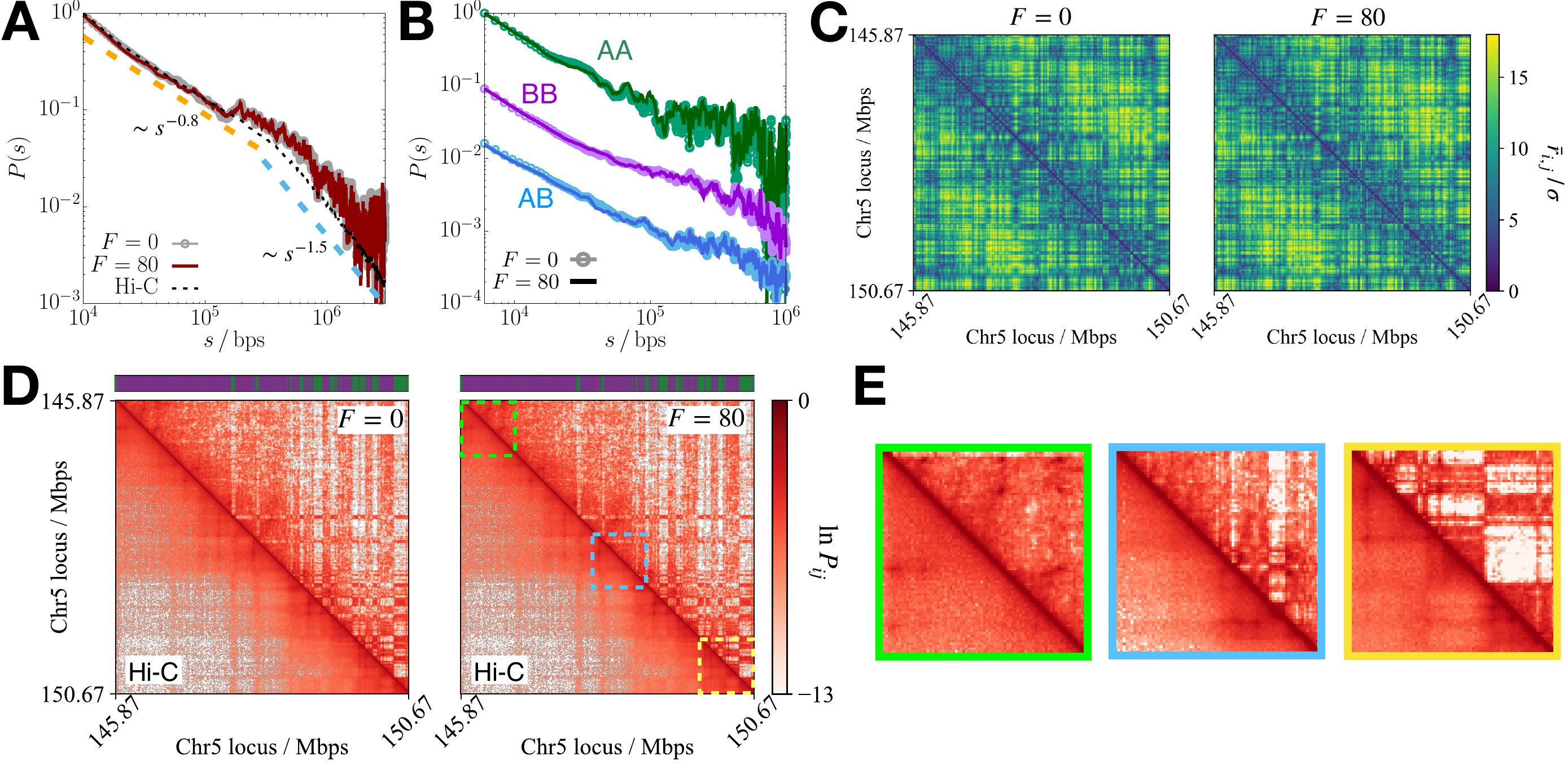}
\caption{Solid-like ordering does not alter the chromatin conformation.
(A) Plot of $P(s)$ (Eq.~\ref{eq:pofs}) at $F=0$ (gray) and $F=80$ (dark red). For comparison, $P(s)$, calculated from the Hi-C data \cite{Rao2014}, is shown in  black dashed line. Two distinct power-law decays are indicated by the orange and blue dashed lines. Each $P(s)$ curve is normalized such that it decays from unity. 
(B) Plots of $P_{\mu\gamma}(s)$ (Eq.~\ref{eq:pofs_type}) at $F=0$ (circles) and $F=80$ (solid line) for A-A (green), B-B (purple), and A-B (blue) pairs. 
$P_\text{AA}(s)$ is normalized such that it decays from unity, whereas $P_\text{BB}(s)$ and $P_\text{AB}(s)$ are shifted from $P_\text{AA}(s)$ in the negative y-direction for a better visualization.
(C) Heat map for the mean distance matrices obtained from the single trajectories at $F=0$ (left) and $F=80$ (right), which start from an  identical configuration.
(D) Comparison between the simulated contact maps (upper triangle) and the Hi-C contact map (lower triangle), where the A/B (green/purple) profile in the model is shown on the top of each map. The predicted contact maps at $F=0$ (left) and $F=80$ (right) are essentially the same. 
(E) Close-up views of three 0.8-Mb regions highlighted by the green, blue, and yellow dashed squares in the righthand contact map in panel D. The green and blue regions show good agreement between the simulated and Hi-C contact maps, whereas there are some quantitative deviations in the yellow region.
Such deviations could arise from the simplicity of the model as well as the finite-size effects, which are in the yellow region that includes a terminus of the polymer with the alternation of locus type.  Note that the green region at the other end, which consists of predominantly B-type locus, shows excellent agreement. The blue region in the middle of the polymer also shows good agreement.
Further adjustment of the interaction parameters ($\epsilon_\text{AA}$, $\epsilon_\text{BB}$, $\epsilon_\text{AB}$) would yield a better agreement, which we did not carry out here.
}
\label{fig:pofs}
\end{figure}

\begin{figure}[h!]
\centering
\includegraphics[width = \textwidth]{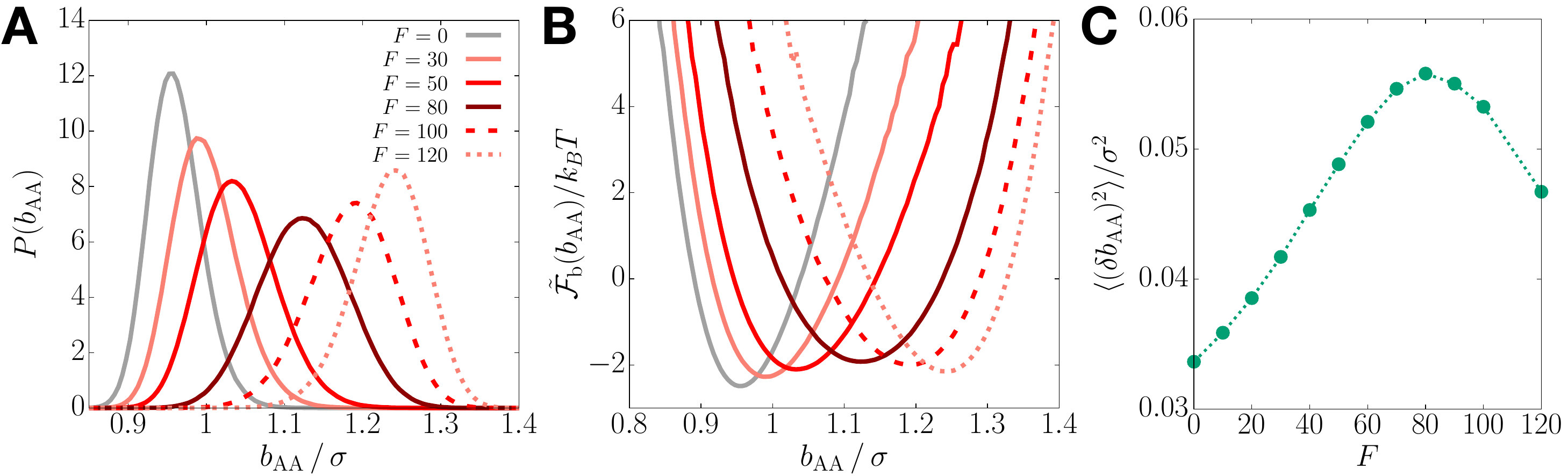}
\caption{Explanation of the ordering mechanism using activity-induced bond length extension. 
(A) Probability distribution of the A-A bond distance for different $F$ values. 
(B) Plots of the effective bond free energy defined by $\tilde{\mathcal{F}}_\text{b}(b_\text{AA})=-k_B T \ln P(b_\text{AA})$.
(C) Bond length fluctuations as a function of $F$. The dotted lines are a guide to the eye. 
}
\label{fig:bond}
\end{figure}

\begin{figure}[h!]
\centering
\includegraphics[width = 5.6 in]{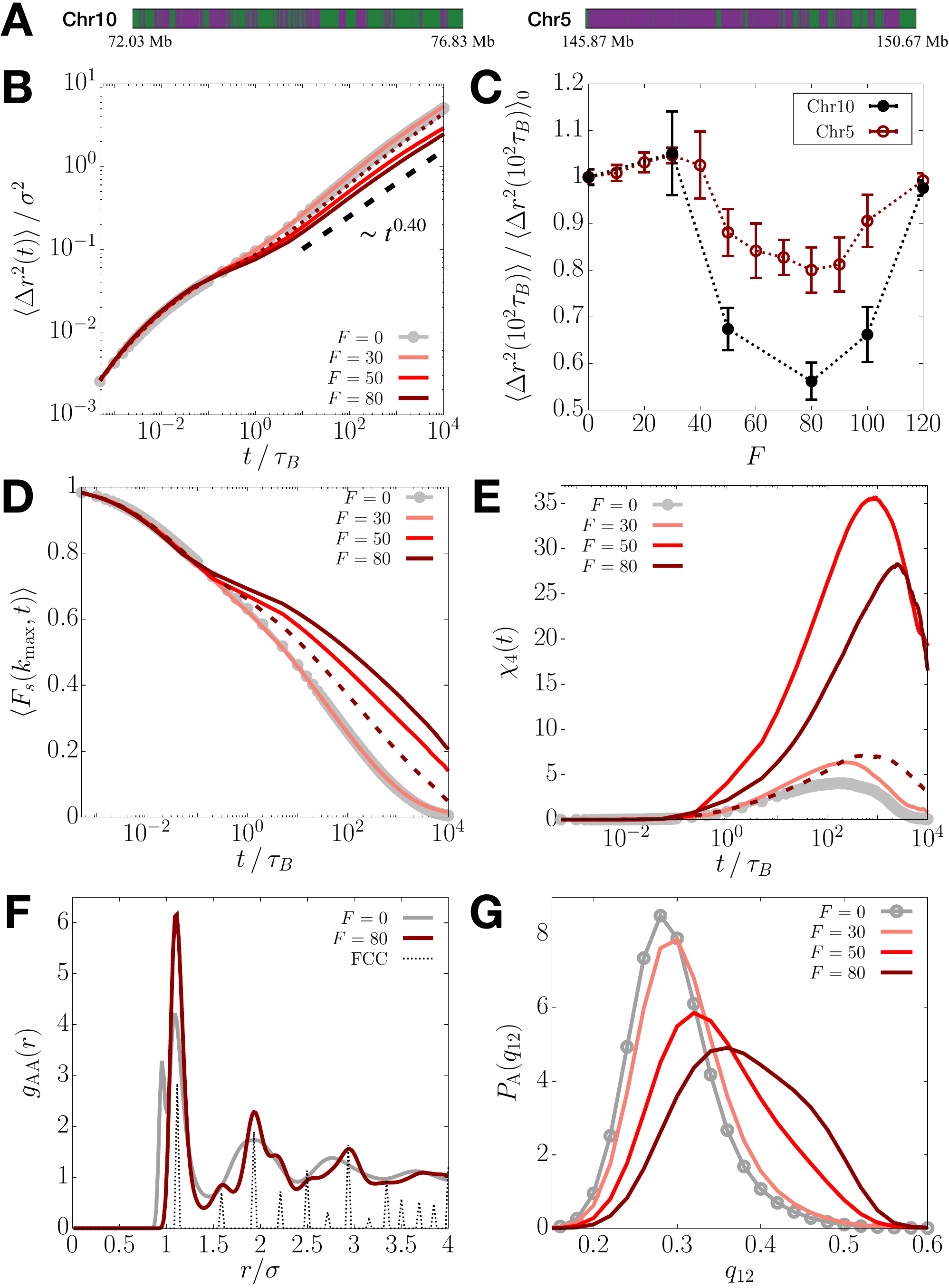}
\caption{Dynamical properties for a 4.8-Mbp segment of Chromosome 10 (Chr10).
(A) Comparison between the A/B (green/purple) profiles for the Chr10 (left; $N_\text{A}/N = 0.58$) and Chr5 (right; $N_\text{A}/N = 0.25$) segments. 
(B) Log-log plots of MSDs for different values of the activity, $F$. For comparison, the MSD curve for Chr5 with $F=80$ is shown by the dotted line.
(C) Ratio of the MSD at $t=100\tau_B$ for a given activity level $F$ to the passive case, where the data for Chr10 (black) and Chr5 (dark-red) are compared. The dotted lines are a guide to the eye. 
(D and E) Plots of $\langle F_s(k_\text{max},t) \rangle$ and $\chi_4(t)$ for different $F$ values. The data for Chr5 with $F=80$ are shown by the dashed lines.
(F) $g_\text{AA}(r)$ at $F=0$ (gray) and $F=80$ (maroon). $g(r)$ for a FCC crystal is shown with the dotted line for comparison.
(G) Plots of $P_\text{A}(q_{12})$ at different $F$. 
}
\label{fig:chr10}
\end{figure}

\begin{figure}[h!]
\centering
\includegraphics[width = 5.6 in]{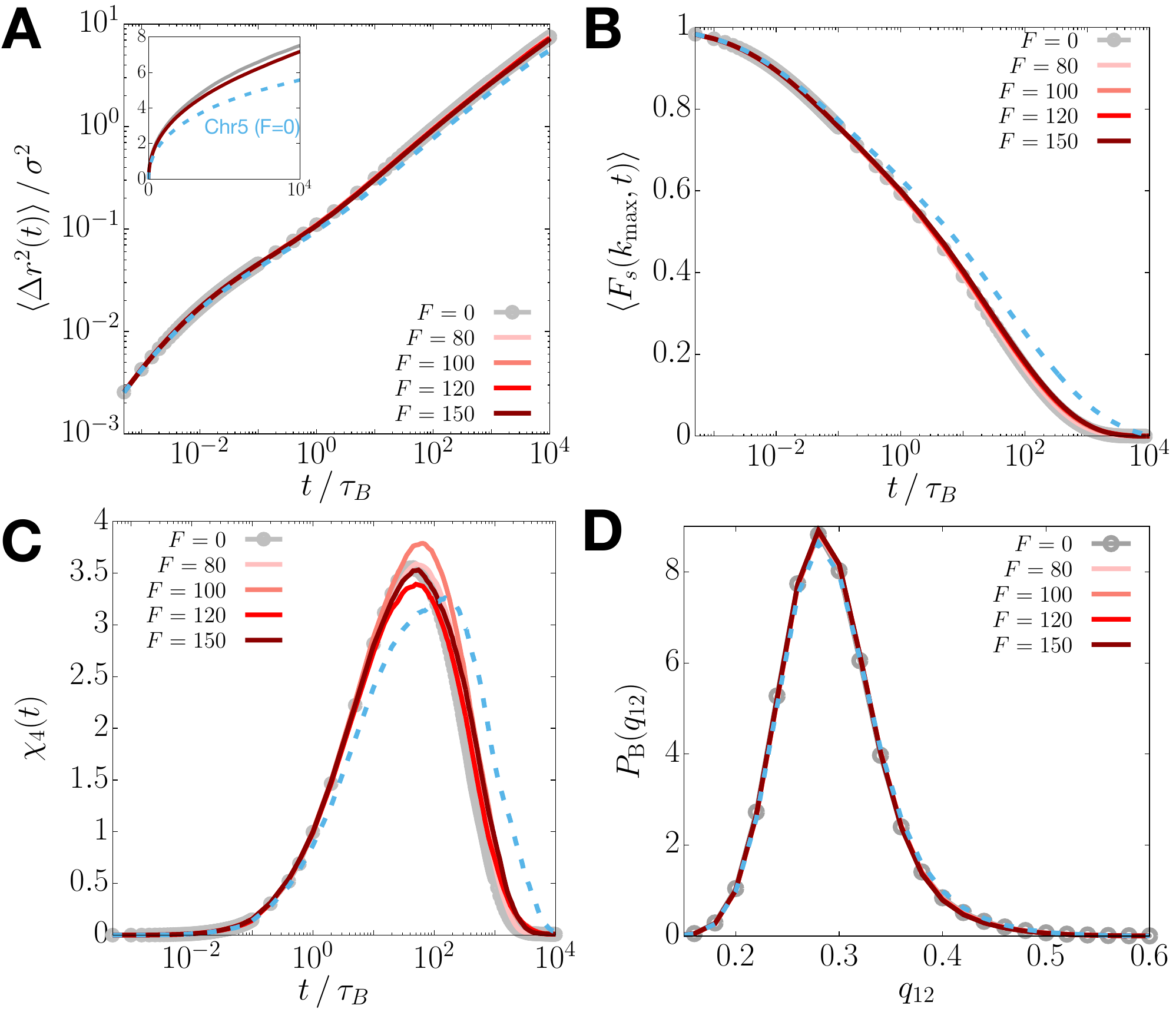}
\caption{A copolymer chain with the same length and the fraction of active loci as Chr5 but randomly shuffled epigenetic sequence does not show dynamical changes and structural ordering upon increasing $F$.
(A) Log-log plots of the MSDs for different values of activity, $F$. For comparison, the MSD curve for the passive Chr 5 is shown by the dashed line. The inset shows the plots in regular scale. 
(B and C) Plots of $\langle F_s(k_\text{max},t) \rangle$ and $\chi_4(t)$ for different $F$ values. The data for the passive Chr5 are shown by the dashed lines.
(D) Probability distributions of $q_{12}$ for B loci at different activity levels, where the dashed line is the distribution for Chr5 with $F=0$, respectively.
}
\label{fig:rand_seq}
\end{figure}

\begin{figure}[h!]
\centering
\includegraphics[width = \textwidth]{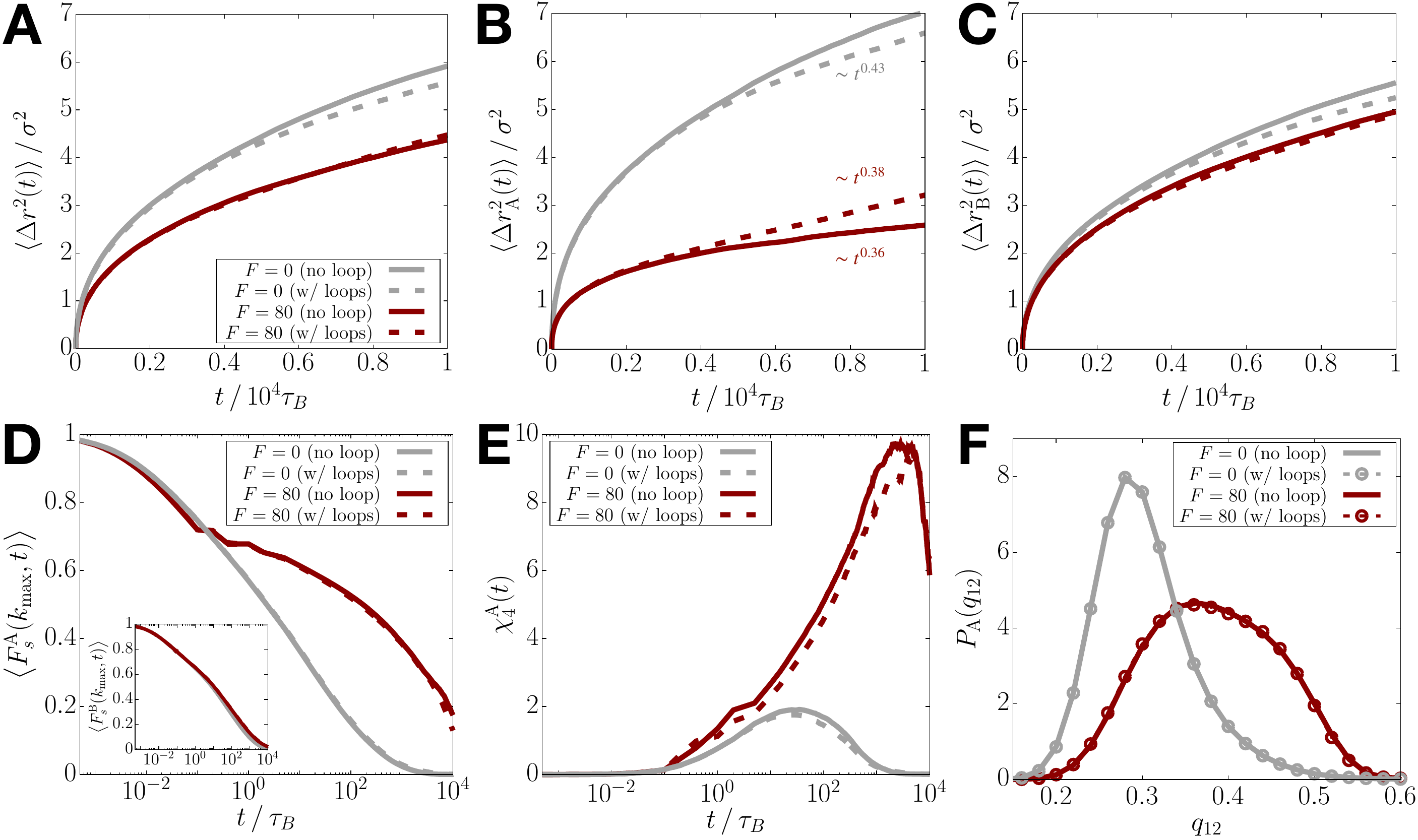}
\caption{Removing the CTCF-mediated loops does not alter the activity-induced suppressed mobilities.
(A)  MSD as a function of time for the Chr5-CCM at $F=0$ (gray) and $F=80$ (dark-red), with (dashed) or without (solid) the CTCF-mediated loops. 
(B and C) Same as panel A, except  the plots show $\langle \Delta r_\mathrm{A}^2(t) \rangle$ and $\langle \Delta r_\mathrm{B}^2(t) \rangle$.
(D) Plots of $\langle F_s^\text{A}(k_\text{max},t) \rangle$ and $\langle F_s^\text{B}(k_\text{max},t) \rangle$ (inset) (Eq.~\ref{eq:fskt}) at $F=0$ (gray) and $F=80$ (dark-red), with (dashed) or without (solid) the loops. 
(E) Same as panel D, except showing the plots of $\chi_4^\mathrm{A}(t)$ (Eq.~\ref{eq:chi4}).
(F) Probability distribution of the BOO parameter for the A loci at $F=0$ (gray) and $F=80$ (dark-red), with (dashed) or without (solid) the loops.
}
\label{fig:loop}
\end{figure}


\begin{figure}[h!]
\centering
\includegraphics[width = 5.6 in]{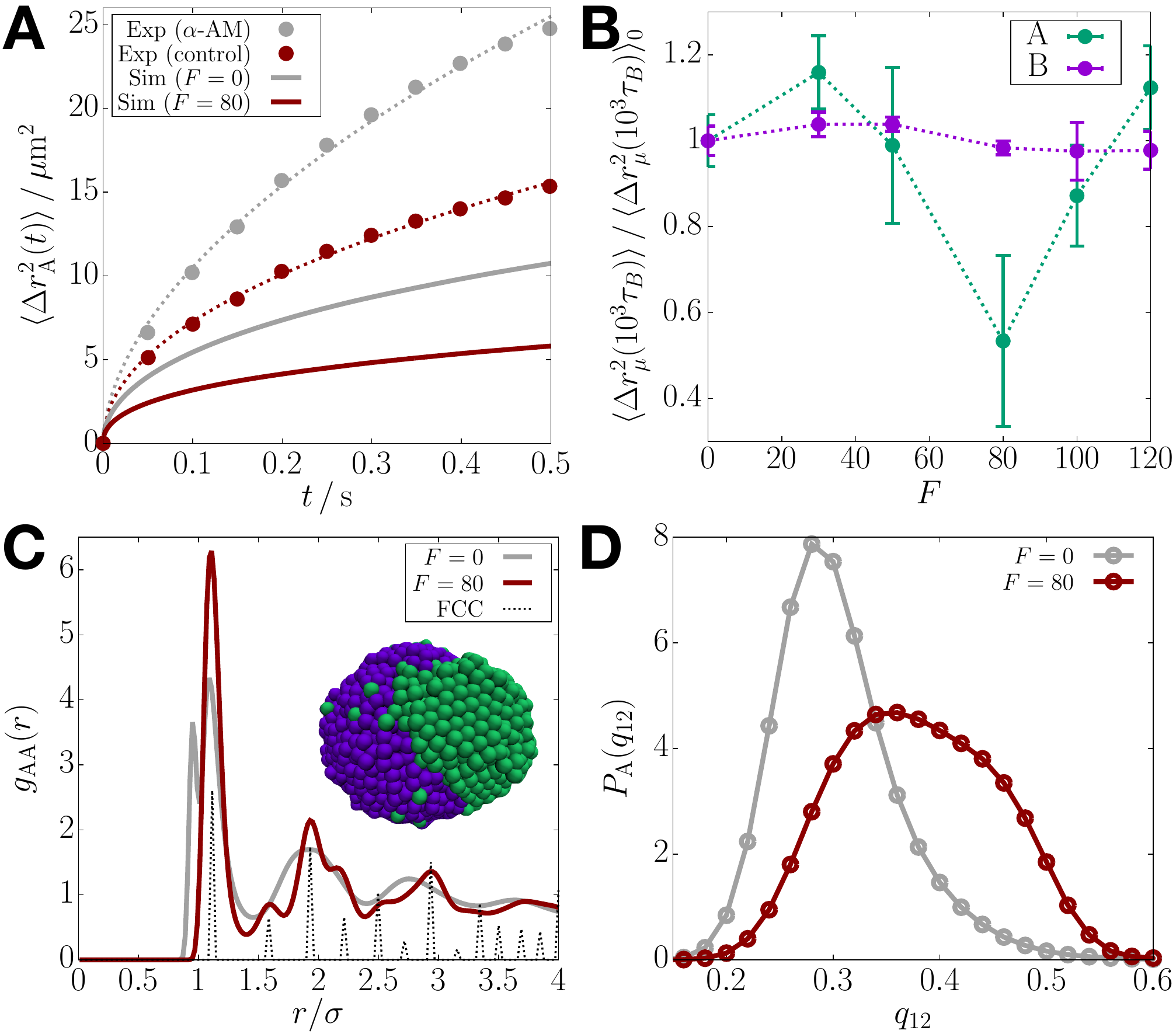}
\caption{Choice of the non-bonding interaction parameters ($\epsilon_\text{AA} < \epsilon_\text{BB}$) does not alter the results of the activity-induced suppressed mobilities and transient ordering.
(A) Plots of $\langle{\Delta r^2_\text{A}}(t)\rangle$ from the CCM simulations for Chr5 with $F=0$ and $F=80$ (solid lines), compared with the euchromatin MSD from the experiment that inhibits transcription using $\alpha$-AM \cite{Nagashima2019} (circles).
(B) Ratio of the MSD at $t=10^3\tau_B$ for a given activity level $F$ to the passive case, where the data for A (green) and B (purple) loci are compared. The dotted lines are a guide to the eye. 
(C) $g_\text{AA}(r)$ at $F=0$ (gray) and $F=80$ (maroon). $g(r)$ for a FCC crystal is shown with the dotted line for comparison. A simulation snapshot for $F=80$ is shown. 
(D) Plots of $P_\text{A}(q_{12})$ at $F=0$ (gray) and $F=80$ (maroon).
}
\label{fig:eps_diff}
\end{figure}

\begin{figure}[h!]
\centering
\includegraphics[width = \textwidth]{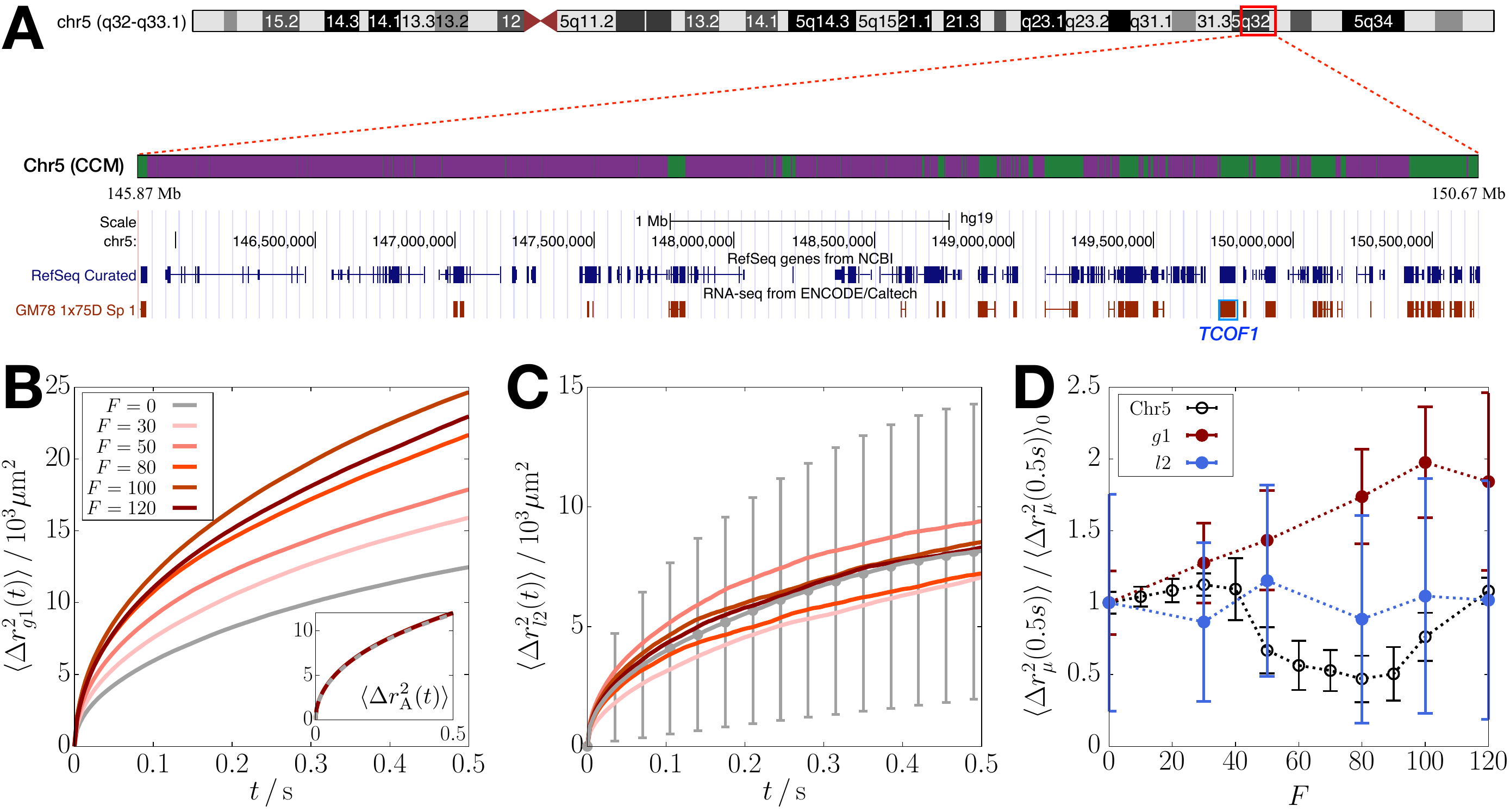}
\caption{Application of the active forces to only one gene region or a few loci enhances mobility.
(A) Screenshot from the UCSC Genome Browser (http://genome.ucsc.edu) \cite{Kent2002} for the region of our interest. The top panel shows the ideogram of Chr5.  The region modeled using the CCM (Chr5: 145.87--150.67 Mb) is highlighted in the red box. In the bottom panel, the first track shows the A/B (green/purple) profile in the CCM. The second track (navy color; right below the scale bar) shows the locations of all the genes in the given region, and the third one (maroon color) indicates the active genes in the GM12878 cell line. The \emph{TCOF1} gene is marked in blue. 
(B)  Time-dependent MSD for the loci corresponding to the \emph{TCOF1} gene body (labeled ``\emph{g1}'') as a function of $F$. The inset shows the MSD for all the A loci at different values of $F$. 
(C) $F$-dependence of the MSD for the first two consecutive A-A bonds in the \emph{TCOF1} gene region (labeled ``\emph{l2}''). The color scheme is the same as that for panel B. 
(D) Ratio of the MSD at $t=0.5\,\mathrm{s}$ at $F$ to the passive case ($F=0$). The data for the single active gene (dark-red) and the two A-A bonds (blue) are compared with the results for the Chr5 region in panel A with all the genes activated (black; \emph{cf}. Fig.~\ref{fig:msd_supp}D). The dotted lines are a guide to the eye. 
}
\label{fig:g1_l2}
\end{figure}

\FloatBarrier

\movie{A simulation trajectory for Chr5 at $F = 80$ and $T = 298\,\mathrm{K}$, corresponding to $2500\tau_\text{B} \approx 1.75\,\mathrm{s}$. Green and purple colors represent the A and B loci, respectively. The light blue spheres indicate the loci with $q_{12}(i) > 0.45$.} 

\movie{A simulation trajectory for Chr5 at $F = 80$ and $T = 3\,\mathrm{K}$ (quenched). The color scheme is the same as in Movie S1.}

\end{document}